\documentstyle[emulateapj,amsmath,epsfig]{article}
\lefthead{Hayashi et al}
\righthead{The structural evolution of substructure}

\def\lsim{\mathrel{\hbox{\rlap{\hbox{\lower4pt\hbox{$\sim$}}}\hbox{$<$}}}}
\def\gsim{\mathrel{\hbox{\rlap{\hbox{\lower4pt\hbox{$\sim$}}}\hbox{$>$}}}}
\def\pz{\phantom{0}}

\newcommand{\kms}{\>{\rm km}\,{\rm s}^{-1}}

\begin{document}

\title{The Structural Evolution of Substructure}

\author{Eric Hayashi\altaffilmark{1,4}, Julio
F. Navarro\altaffilmark{1,5}, James E. Taylor\altaffilmark{1,2}, \\ Joachim
Stadel\altaffilmark{1} \& Thomas Quinn\altaffilmark{3}}

\affil{$^1$Department of Physics and Astronomy, University of Victoria, 
 Victoria, BC, V8P 1A1, Canada}

\affil{$^2$Department of Physics, Denys Wilkinson Building,
Keble Road, University of Oxford, Oxford OX1 3RH, UK}

\affil{$^3$ Department of Astronomy, University of Washington, Seattle, WA
98195, USA.}

%%%%%%%%%%%%%%
% Additional affiliations
%%%%%%%%%%%%%%

\altaffiltext{4}{e-mail: ehayashi@uvastro.phys.uvic.ca}
\altaffiltext{5}{Scholar of the Canadian Institute for Advanced Research}

%\setcounter{footnote}{5}

%%%%%%%%%%%%%%%
% Abstract
%%%%%%%%%%%%%%%

\begin{abstract}
We investigate the evolution of substructure in cold dark matter halos using
N-body simulations of tidal stripping of substructure halos within a static host
potential. We find that halos modeled following the Navarro, Frenk \& White
(NFW) mass profile lose mass continuously due to tides from the massive host,
leading to the total disruption of satellite halos with small tidal radii.
Although mass is predominantly stripped from the outer regions, tidal heating
also causes the halo to expand and the central density to decrease after each
pericentric passage, when mass loss preferentially occurs. As a result, simple
models based on the tidal-limit approximation underestimate significantly the
tidal mass loss over several orbits. The equilibrium structure of stripped NFW
halos depends mainly on the fraction of mass lost, and can be expressed in terms
of a simple correction to the original NFW profile. We apply these results to
substructure in the Milky Way, and conclude that the dark matter halos
surrounding its dwarf spheroidal (dSph) satellites have circular velocity curves
that peak well beyond the luminous radius at velocities significantly higher
than expected from the stellar line-of-sight velocity dispersion. Our modeling
suggests that the true tidal radii of dSphs lie well beyond the putative tidal
cutoff observed in the surface brightness profile, suggesting that the latter
are not really tidal in origin but rather features in the light profile of
limited dynamical relevance.  For Draco, in particular, our modeling implies
that its tidal radius is much larger than derived by Irwin \& Hatzidimitriou
(1995), lending support to the interpretation of recent Sloan survey data by
Odenkirchen et al.~(2001). Similarly, our model suggests that Carina's halo has
a peak circular velocity of order $\sim 55 \kms$, which may help explain how
this small galaxy has managed to retain enough gas to undergo several bursts of
star formation. Our results imply a close correspondence between the most
massive substructure halos expected in a CDM universe and the known satellites
of the Milky Way, and suggest that only substructure halos with peak circular
velocities below $35$ km s$^{-1}$ lack readily detectable luminous counterparts.
\end{abstract}

\keywords{N-body simulations; galaxies: evolution, structure; cosmology: dark
matter}

\section{Introduction}
\label{sec:intro}

Over the past two decades, cosmological models based on the assumption that the
Universe is gravitationally dominated by Cold Dark Matter (CDM) have evolved
into the prevailing paradigm for interpreting the formation and evolution of
structure in the Universe (Blumenthal et al.~1984, Davis et al.~1985, Bahcall et
al.~1999). In this model, structure develops as primordial density fluctuations
with scale-invariant power spectra grow to form virialized structures. Due to
the negligible thermal velocity of CDM, fluctuations survive the early universe on
all scales, and structure develops from the bottom up as small, dense CDM clumps
collapse at very early times and subsequently undergo a series of mergers that
result in the hierarchical formation of large, massive dark matter halos.

These halos constitute the hosts of all galaxy systems, from individual galaxies
to galaxy groups and clusters. In a pioneering paper, White \& Rees (1978) noted
that mergers of dark matter clumps should very efficiently disrupt the
progenitors and lead to smooth remnants where most of the mass is in a single,
monolithic structure. Early cosmological N-body simulations showed that the
remnants of the merger hierarchy were featureless systems with little
substructure, and appeared to confirm this expectation. The survival of
individual galaxies in groups and clusters was thus linked to the ability of
baryons to cool and condense into much more tightly bound structures than their
surrounding dark halos, allowing them to survive as self-bound entities the
assembly of galaxy clusters.

Recently, the development of efficient algorithms and the advent of
massively-parallel computers have led to simulations with dramatically improved
force and mass resolution. These have revealed a wealth of substructure in
virialized dark matter halos, where none was apparent at lower resolution
(Ghigna et al.~1998, Klypin et al.~1999b, hereafter K99b, Springel et
al.~2000). Substructure appears to extend down to the smallest mass scales
reliably resolved, and several hundred dark matter satellites are expected
within a few hundred kpc of the Milky Way.  Most of these substructure halos,
hereafter `subhalos', are of low mass. Indeed, as a whole, subhalos contribute
only about $10\%$ of the mass of a virialized halo, and most of the mass is in a
smooth monolithic structure, as envisioned by White \& Rees (1978).

These results imply that CDM halos are quite resilient to tidal disruption,
since they may lose a large fraction of their mass without being completely
disrupted. This resilience stems from the large concentration of their mass
distribution, which is in turn intimately linked to the negligible thermal speed
of CDM particles. Indeed, models where dark matter is `hot' or `warm' rather
than cold lead to the formation of halos with much reduced substructure (Colin
et al.~2000, Bode, Ostriker \& Turok 2001, Avila-Reese et al.~2001, Dalcanton \&
Hogan 2001, Knebe et al.~2002).

The dependence of substructure on the nature of dark matter makes it a
discriminating tool between competing cosmogonical models. For example, Klypin
et al.~(1999a, hereafter K99a) and Moore et al.~(1999, hereafter M99) have
pointed out that the number of low mass subhalos identified in CDM simulations
of a Milky Way-sized galaxy halo exceeds the number of observed Galactic
satellites by a factor of $10$ to $100$. This observation has fueled speculation
that it may be necessary to modify radically the CDM paradigm on small scales in
order to account for the scarcity of luminous satellites in the vicinity of the
Milky Way.

Possible modifications include a finite interaction cross-section for dark
matter particles (Spergel \& Steinhardt 2000), or a finite `temperature' for
dark matter, as in the warm dark matter scenario (Dalcanton \& Hogan 2001).
Whether or not such radical modifications are actually necessary is still being
debated, although it remains possible that the apparent `decoupling' between
dark and luminous matter in low-mass halos may be a natural consequence of the
effect of stellar feedback on these scales (White \& Rees 1978, Kauffmann, White
\& Guiderdoni 1993, Cole et al 1994), or perhaps of the heating effects of an
ionizing UV background (Efstathiou 1992, Bullock, Kravtsov \& Weinberg 2000,
Somerville 2001, Benson et al.~2001).

So far, most of the effort devoted to studying the properties and evolution of
substructure have focused on a few cosmological simulations with ultra-high
numerical resolution, such as those performed by Moore et al.~(1998), Ghigna et
al.~(2000) and Klypin et al.~(2001). Such studies are hindered by the small
fraction ($\sim 10\%$) of mass that remains attached to self-bound subhalos in a
typical virialized system.  As a result, even in the best simulations a typical
subhalo is made up of little more than $\sim 10^3$ particles and is particularly
vulnerable to numerical artifacts affecting small-N systems near the resolution
limit. In addition, the complexity of the cosmological context, with its
continuous mergers and rapidly changing potential wells, makes it difficult to
isolate the mechanisms responsible for driving the evolution of
substructure. Many important questions thus remain unanswered: for example, (i)
are substructure halos fully disrupted by the tidal field of the host system?;
(ii) how does the internal structure of substructure halos evolve as they are
tidally stripped?; (iii) how do numerical limitations affect the mass function
of subhalos?

Addressing these questions requires much better resolution than is achievable in
current cosmological simulations. We have therefore decided to investigate the
evolution of subhalos by performing high-resolution simulations of the evolution
of realistic dark halo models within the gravitational potential of a much
larger system. The model we adopt assumes a spherically-symmetric and 
time-independent host potential, but accounts fully for the self-interaction
between particles in the subhalo. Our models are thus useful to study in detail
the process of tidal heating and stripping by the main potential but neglect the
complicating effects of dynamical friction, of harassment between subhalos, and
of severe departures from spherical symmetry in the host. Despite these caveats,
our simulations can be usefully applied to the evolution of systems much less
massive than the host, as is the case for individual galaxies in a rich cluster
or for satellites orbiting a bright galaxy.

The outline of this paper is as follows.  In Section \ref{sec:halo} we present
our halo model and discuss the properties of equilibrium N-body models evolved
in isolation. Section \ref{sec:tidaldis} describes our simulations of satellite
halos on bound orbits in the gravitational potential of a static host.  Section
\ref{sec:modeling} compares the results of these simulations with the
predictions of simple models based on the tidal limit and on the impulse
approximation. Section \ref{sec:stripprof} examines the changes in the internal
structure of tidally stripped halos and Section \ref{sec:mwsub} explores its
implications for the properties of dark halos surrounding the dwarf spheroidal
companions of the Milky Way. We summarize our main conclusions in Section
\ref{sec:concs}.

\section{The Halo Model}
\label{sec:halo}

The survival of a dark matter halo in the presence of strong tides depends
sensitively on its internal structure. It is therefore important to choose a
halo model that represents as faithfully as possible the structure of dark
matter clumps expected to form in the cosmological model to be investigated. The
density profile of cold dark matter halos has been the subject of extensive
numerical studies by numerous authors (Frenk et al.~1988, Dubinski \& Carlberg
1991, Warren et al.~1992, Navarro, Frenk \& White 1996, 1997, Fukushige \&
Makino 1997, Moore et al.~1998), all of which agree on two basic results: (i)
the density profile of CDM halos is shallower than isothermal near the center
and steepens gradually outwards, becoming steeper than isothermal in the outer
regions; and (ii) there is no well defined value for the central density of the
dark matter, which can in principle climb to arbitrarily large values near the
center.

Navarro, Frenk \& White (1996, 1997, hereafter NFW) propose a simple formula
that synthesizes this result and that provides a good description of the
spherically-averaged density profile of virialized CDM halos,
\begin{equation}
\label{eq:nfw}
\rho_{\rm NFW}(r) =\frac{\rho_s}{r/r_s(1 + r/r_s)^2},
\end{equation}
where $r_s$ is a characteristic scale radius. Although there is some
disagreement regarding how accurately this simple formula describes the latest
high-resolution simulations (Fukushige \& Makino 1997, Moore et al.~1998, Ghigna
et al.~2000, Klypin et al.~2001), the enclosed mass profile it predicts seems to
agree well with high-resolution simulations of galaxy-sized systems (Navarro
2001, Power et al.~2002).  In view of this, and given its simplicity, we adopt
this model for our study.

\begin{figure}
\centerline{\epsfbox{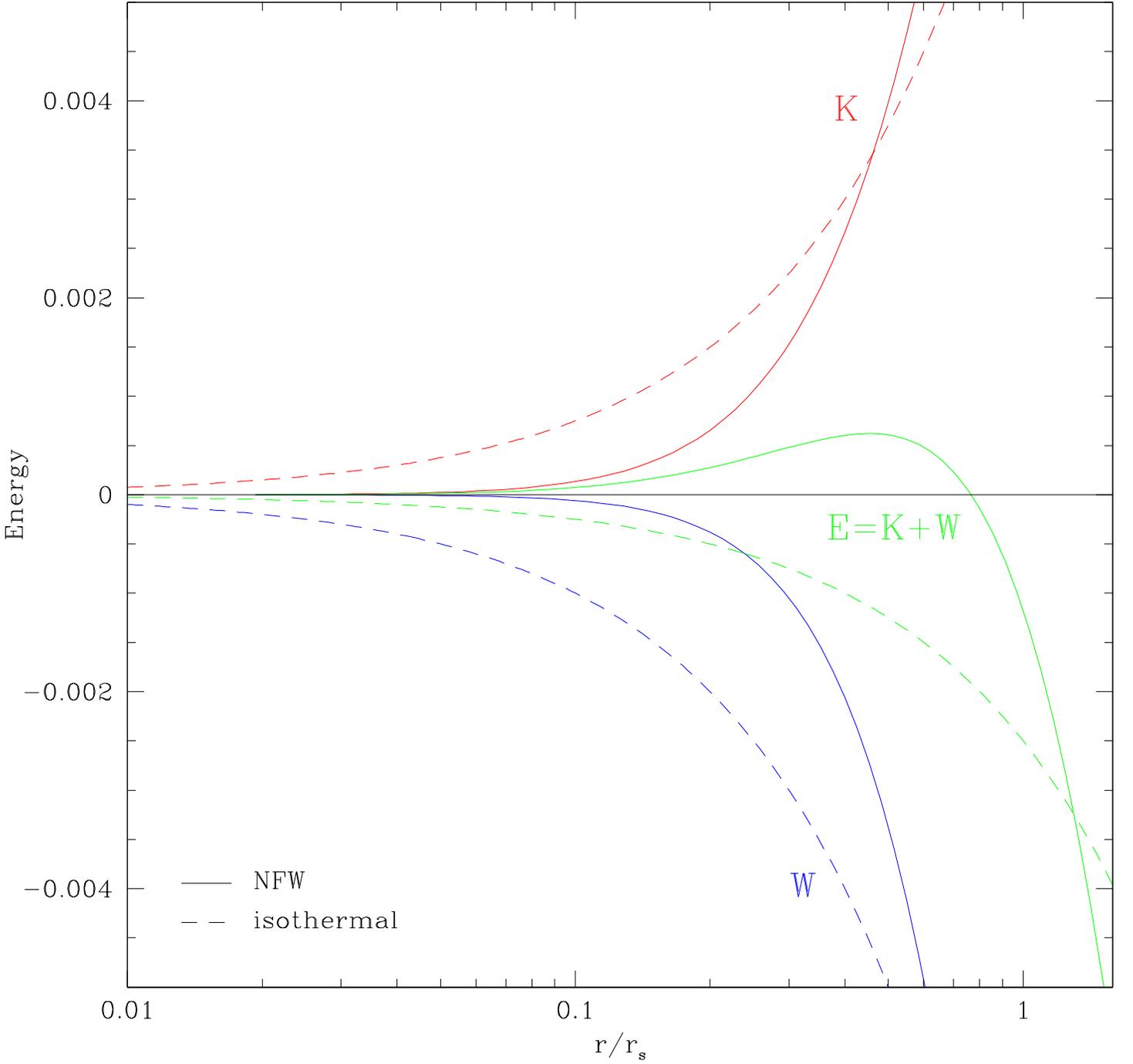}}
\figcaption{{Kinetic ($K$), potential ($W$) and total ($E = K+W$) energy
profiles for NFW and isothermal distributions truncated at radius $r$, in
arbitrary units.  Note that the NFW halo (solid lines) has positive total energy
if truncated at radii smaller than $r_{\rm bind} \simeq 0.77\, r_s$, whereas the
isothermal halo (dashed lines) remains bound at all radii.\label{fig:nfwiso}}}
\end{figure}

One important feature of models such as NFW is that, when truncated, the total
energy of the remaining system is not necessarily negative. This is shown in
Figure~\ref{fig:nfwiso}, where we plot the kinetic and potential energy of an
isotropic NFW halo truncated `instantaneously' at radius $r$ (solid lines). The
total energy of the truncated system is negative only when the truncation radius
is larger than a `minimum binding radius', $r_{\rm bind} \simeq 0.77 \, r_s$. This
suggests that NFW models tidally limited to radii comparable to $r_{\rm bind}$
(or, equivalently, that are stripped of more than $\sim 91\%$ of their mass)
may be totally dispersed by tides.

The dashed lines in Figure \ref{fig:nfwiso} show that this is not the case for a
truncated singular isothermal sphere, which remains self-bound regardless of the
truncation radius. As discussed by Moore, Katz, \& Lake (1996), halos with
singular isothermal density profiles ``should always survive at some level,''
but halos with shallower inner density profiles like NFW may disrupt unless a
dissipational component is included. Our simulations are designed to study the
mass loss timescale for halos orbiting within more massive systems, as well as
the conditions under which disruption by tides may occur.

\subsection{N-body Realizations of NFW Halos}
\label{sec:isolated}

Our procedure for initializing isotropic N-body realizations of NFW models
follows closely the prescriptions of Hernquist (1993). Since NFW models have
infinite mass, it is necessary to truncate them at some fiducial radius, which
we take to be $r_{\rm cut}=10 \, r_s$.  With this choice, the half-mass radius
of the model is $r_{h} \simeq 3.6 \, r_s$. We choose the spline softening length
of our runs, $\epsilon$, according to the scaling suggested by van Kampen
(2000), so that $\epsilon= 0.77 \, r_{h} \, N^{-1/3}$, where $N$ is the total
number of particles. With this scaling, $\epsilon=0.06 \, r_s$ and
$\epsilon=0.19 \, r_s$ for realizations with $N=10^5$ and $N=3000$,
respectively.{\footnote{Dehnen (2001) suggests an alternative scaling for the
optimal softening length, $\epsilon \propto N^{1/5}$ for compact softening
kernels.  With this scaling, $\epsilon=0.12 \, r_s$ and $\epsilon=0.21 \, r_s$
for halo realizations with $N=10^5$ and $N=3000$, respectively.}}

Unless otherwise specified, we shall adopt units where the gravitational
constant, $G$, the scale radius, $r_s$, and the total mass of the satellite
halo, $m_{\rm sat}$, are all unity.  With this choice, the crossing time of the
system is $t_{\rm cross}=\sqrt{r_{\rm cut}^3/G m_{\rm sat}}=31.6$ and the
circular orbit timescale at the half-mass radius is $t_{\rm circ}(r_h)=2\pi
r_h/V_c(r_h)=60.7$. All simulations have been carried out using Stadel \&
Quinn's multi-stepping, parallel treecode {\tt PKDGRAV} (Stadel 2001). We use a
kick-drift-kick leapfrog integrator with adaptive, individual timesteps chosen
according to the gravitational potential and the acceleration, $\Delta t_i=0.03
\, \sqrt{\Phi_i}/a_i$. Timesteps in isolated runs typically range between $2.5
\times 10^{-4}$ and $4 \times 10^{-3} \, t_{\rm cross}$, with a median value of
$10^{-3} \, t_{\rm cross}$.

\subsection{Secular Evolution of N-body Halo Models}
\label{ssec:secev}

Isolated N-body halo models evolve away from their equilibrium configuration due
to discreteness effects associated with the finite number of particles. The main
mechanism driving this secular evolution is encounters between particles, which
lead to energy exchanges between different parts of the system and,
consequently, to gradual modifications in the structure of the halo.

It is instructive to consider the specific kinetic energy of particles at
various radii throughout the system, since collisions will naturally tend to
drive the system towards equipartition. The velocity dispersion in isotropic NFW
halos rises from the center outwards, reaches a maximum roughly at $r_s$, and
decreases further out. This behaviour is mimicked by the circular velocity,
which peaks at $r_{\rm max}=2.16 \, r_s$. This radial dependence of the `temperature'
of the system determines to a large extent the main qualitative features of the
relaxation-driven evolution of an N-body NFW model.

\begin{figure}
\centerline{\epsfbox{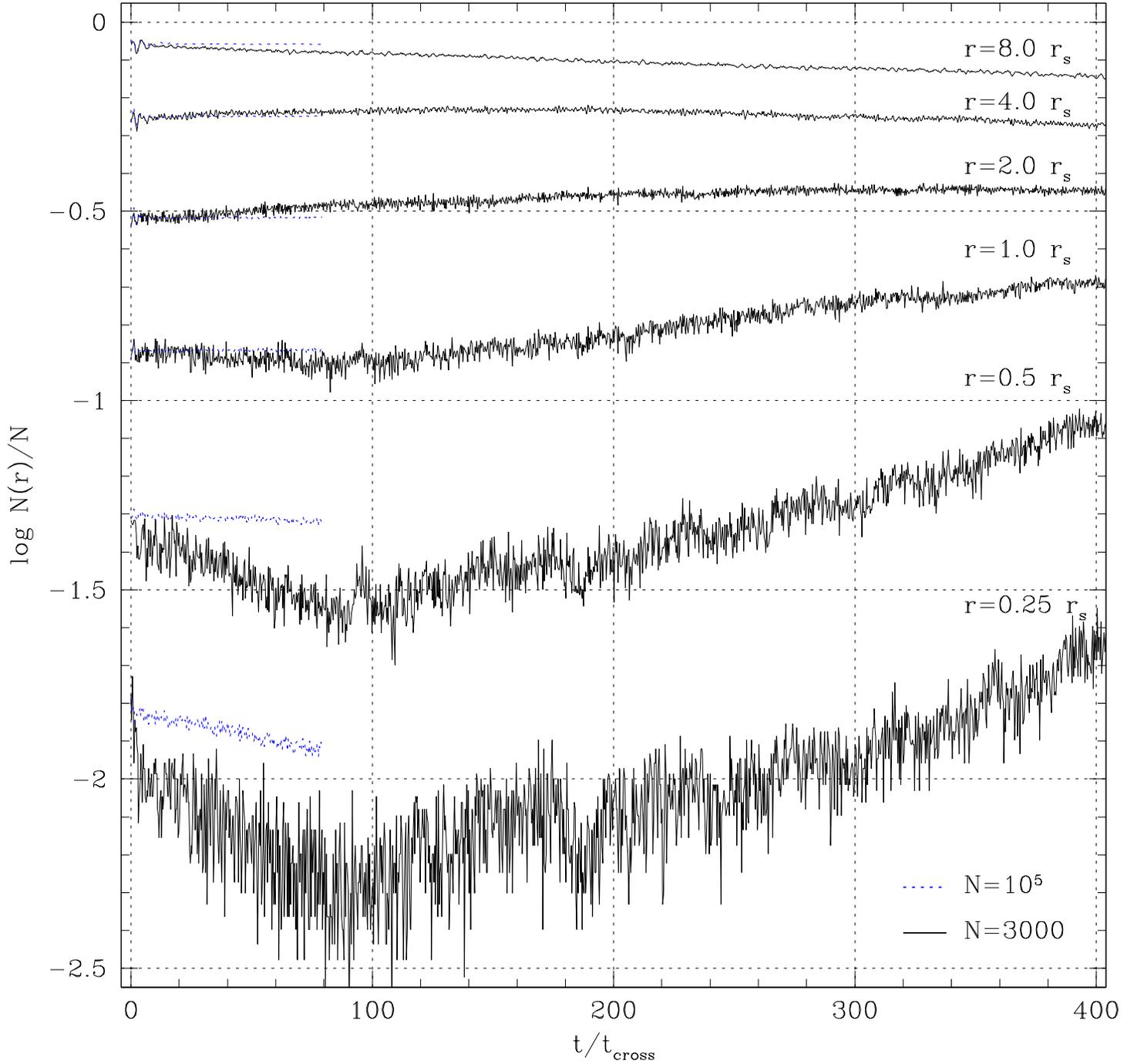}} 
\figcaption{From top to bottom, the mass fraction enclosed within $8.0, 4.0,
2.0, 1.0, 0.5$ and $0.25 \, r_s$ for $N=3000$ (solid) and $N=10^5$ (dotted)
isolated NFW halo models as a function of time.\label{fig:mrad3k}}
\end{figure}

The solid (dotted) curves in Figure~\ref{fig:mrad3k} show the mass within
various radii for an $N=3000$ ($N=10^5$) model evolved for up to $400$ ($80$)
crossing times. Two different evolutionary stages can be seen over this very
long timescale. At first ($t \lsim 100\, t_{\rm cross}$), the region within one
scale radius (which is originally cooler than its surroundings) is heated
through collisions with faster-moving particles and gradually expands to
accommodate the energy gained. This is shown in Figure~\ref{fig:mrad3k}, where
the mass enclosed within radii smaller than $r_s$ is seen to decrease
systematically for $t < 100\, t_{\rm cross}$.  A similar process appears to
operate well beyond $r_s$, which is seen to gain energy (i.e., to become less
dense) at the expense of the hotter, inner regions, albeit on a much longer
timescale (see, e.g., the curve labelled $8.0\, r_s$ in
Figure~\ref{fig:mrad3k}).  To conserve energy, intermediate regions of the
system become increasingly bound and dense: the mass within $2\, r_s$ is seen to
increase steadily throughout the evolution. We have explicitly checked that the
evolution depicted in Figure~\ref{fig:mrad3k} is not affected by integration
errors; our conservative timestepping choice leads to energy conservation better
than $0.32\%$ even after $400$ crossing times.  We have also ruled out
non-equilibrium initial conditions as the cause of the observed evolution.
After some initial fluctuations within the first few crossing times, the ratio
of kinetic to potential energy of the halo remains constant to within less than
$2\%$ for the duration of the simulation.

For $N=3000$, energy transfer into the inner core shuts off at $t\approx 100\,
t_{\rm cross}$, when the velocity dispersion in the inner regions (`core') has
been raised to roughly match that at $r_s$.  The second stage of the process
begins then, in which the core contracts as more and more particles are kicked
into highly energetic orbits. This `core collapse' stage, familiar from globular
cluster evolution calculations, leads gradually to highly concentrated systems:
after $400$ crossing times the density within $0.25\, r_s$ has gone up by almost
a factor of $3$ and the mass within $r_s$ has nearly doubled.

\begin{figure}
\centerline{\epsfbox{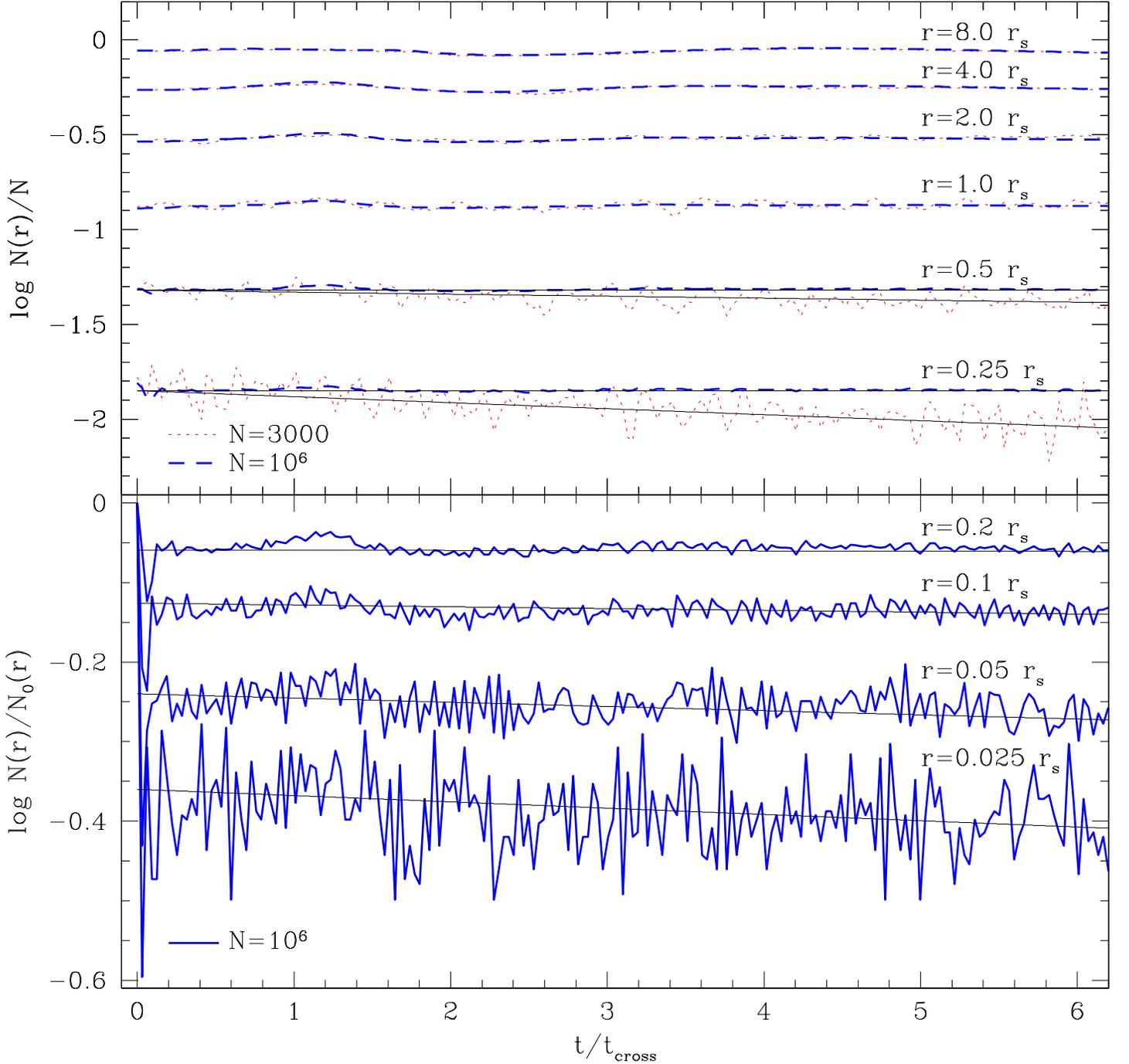}} 
\figcaption{
{\it Top panel:} From top to bottom, the mass within $8.0, 4.0, 2.0, 1.0, 0.5$
and $0.25 \, r_s$ for $N=10^6$ (solid blue) and $N=3000$ (dotted red) NFW halo
models as a function of time.  Solid black lines show the decline in mass within
$r=0.5$ and $0.25\, r_s$ expected from the local evaporation timescale of each
halo given by eq.~\ref{eq:tevap}. {\it Bottom panel:} From top to bottom, time
evolution of the mass within $0.2, 0.1, 0.05$ and $0.025 \, r_s$ for the
$N=10^6$ halo, normalized to the initial mass within each radius $N_0(r)$.
Solid black lines show fits with exponential decay times equal to the
evaporation timescale appropriate for each radius.
\label{fig:mrad}}
\end{figure}

On what timescale does the inner mass profile evolve?  The bottom panel in
Figure~\ref{fig:mrad} shows the mass within $0.025$, $0.05$, $0.1$, and $0.2 \,
r_s$, respectively, for a run with $N=10^6$ particles. The mass within these
radii (shown in units of the theoretical mass within each radius $N_0(r)$)
decreases continuously and at rates that increase gradually nearer the
center. We measure the timescale of this evolution from fits of the form,
$N(t)=N_0 \, {\rm e}^{-t/t_0}$, shown as straight solid lines in
Figure~\ref{fig:mrad}. These timescales are listed in Table~\ref{tab:1M}, and
exceed by about two orders of magnitude the local collisional relaxation
timescale,
\begin{equation}
t_{\rm relax}(r) \approx \frac{0.1 \, N(r)}{\ln N(r)}
\left({{\bar \rho}(r) \over {\bar \rho}(r_{\rm cut})}\right)^{-1/2} t_{\rm circ}(r_{\rm
cut}).
\label{eq:trelax}
\end{equation}
Indeed, Table~\ref{tab:1M} shows that mass is evacuated from the center at rates
consistent with the {\it evaporation} timescale, 
\begin{equation}
t_{\rm evap}(r)\sim 136 \, t_{\rm relax}(r)
\label{eq:tevap}
\end{equation}
(Binney \& Tremaine 1987). This is not entirely unexpected; individual particle
energies {\it do} change on the relaxation timescale, but it is the slower
process of evaporation that leads to appreciable systematic changes in the mass
profile of halo models.  The straight lines in the top panel of
Figure~\ref{fig:mrad} confirm that the same considerations also describe quite
well the evolution of the inner mass profile in the $3000$-particle runs: these
are {\it not} fits to the data, but rather show the decline in mass expected
from the local evaporation timescale computed from eq.~\ref{eq:tevap}.

We note that the secular evolution of spherically-symmetric N-body systems has
been investigated previously by Quinlan (1996) using Fokker-Planck
methods. Quinlan's study indicates that collisional effects start modifying the
mass profile on the local relaxation timescale, whereas ours suggest that, once
started, the mass profile gets modified at a rate controlled by the local
evaporation timescale. Quantitatively, our results are in reasonably good
agreement with Quinlan's work. For example, scaling his results to our NFW
models with $10^5$ particles, we would expect the mean density within a radius
containing roughly 1\% of the total mass to decrease by about 50\% over a
timescale equivalent to $100 t_{cross}$.  This is in reasonable agreement with
the results presented in Figure 2, where the mean density within a fixed radius
$r=0.25~r_s$ (initially containing 1.6\% of the total mass), is seen to decrease
by about 30\% over a similar timescale.

In summary, we expect N-body models of NFW halos to deviate significantly from
their original structure on the local evaporation timescale. Halos tend to
develop an isothermal ``core'' of size $r \sim r_s$ on a timescale comparable to
$t_{\rm evap}(r_s)$. Over longer periods, the central regions are expected to
become significantly denser, and thus more resilient to tides. This affects
primarily small-$N$ systems, where relaxation timescales are short.

We can estimate the importance of this process in cosmological N-body
simulations by identifying the outer radius, $r_{\rm cut}$, with the `virial
radius', $r_{200}$, where the mean density contrast{\footnote{We use the term
`density contrast' to denote densities expressed in units of the critical
density for closure, $\rho_{\rm crit}=3H^2/8\pi G$. We express the present value
of Hubble's constant as $H(z=0)=H_0=100\, h$ km s$^{-1}$ Mpc$^{-1}$}} is
200. With this choice, $t_{\rm circ}(r_{\rm cut})$ is comparable to the age of
the universe at the time a halo is identified. Thus, halos formed at $z=10$,
$5$, and $1$, would have evolved (if kept isolated) for $570$, $218$, and $26$
crossing times by $z=0$, respectively. (We have assumed a $\Lambda$CDM cosmogony
with $\Omega_0=0.3$, $\Omega_{\Lambda}=0.7$, $h=0.65$).  According to
eq.~\ref{eq:tevap}, the age of such systems at $z=0$ will match their half-mass
evaporation timescales for $N=200, 55,$ and $N < 1$, respectively. We conclude
that it is unlikely that relaxation effects will introduce a significant bias in
the evolution of substructure halos with $N > 200$. The survival of halos with
$N\lsim 200$ collapsing at $z < 10$, on the other hand, should be carefully
monitored to rule out the possibility that they may be affected by relaxation
effects.

\section{Tidal Stripping of Subhalos}
\label{sec:tidaldis}
\subsection{The Numerical Setup}
\label{sec:sims}

We use the NFW halo models described in the previous section to investigate
quantitatively the tidal stripping of `satellite' halos orbiting within a much
more massive host system. To this effect, after allowing an NFW model to relax
for $10$ crossing times, we place it in orbit within a static,
spherically-symmetric NFW potential with scale radius $R_s=10\, r_s$, and total
mass, $M_{\rm host}$, within $10\, R_s$. Our simulations thus neglect the
effects of dynamical friction and the reaction of the host to the satellite, but
given the vast difference in mass and size of the two systems these are expected
to contribute only minor corrections to our results. For example, applying the
approximate formula of Colpi, Mayer \& Governato (1999) we find that the
dynamical friction timescale for an orbit with $r_{\rm ap} = 10\, R_s$ and
$r_{\rm per} = 1.5\, R_s$ exceeds the orbital timescale by up to a factor of
$\sim 20$.

Note that due to the scale-free nature of gravity, the physical scaling of our
systems is arbitrary. For example, if we take $r_s \approx 10~{\rm kpc}$ and
$m_{\rm sat} \approx 10^{12}\, M_{\odot}$ the satellite halo would represent a
Milky Way-sized galaxy, and the host potential would corresponds to that of a
rich galaxy cluster.  Choosing $r_s \approx 1$ kpc and $m_{\rm sat} \approx
10^{9}\, M_{\odot}$, on the other hand, would correspond to the case of a dwarf
galaxy orbiting within the potential of a Milky Way-sized halo.

We have performed a series of simulations varying the orbit, the mass resolution
of the satellite, and the mass of the host relative to that of the satellite.
In particular, we have followed the evolution of $N=3000$ and $N=10^5$ halos on
three different eccentric orbits, all of which have the same apocentric radius,
$r_{\rm ap} = 10\,R_s$, but different pericenters, $r_{\rm per} = 1.5, 3.0,$ and
$6.0\,R_s$, respectively.  Note that the median apocenter to pericenter ratio of
subhalos in cosmological N-body simulations is between 5:1 and 6:1 (Ghigna et
al.~1998, Tormen, Diaferio, \& Syer 1998), comparable to some of our eccentric
orbits (see Table~\ref{tab:orbparams}).  Several 3000-particle satellites were
also placed on circular orbits in this potential, with orbital radii ranging
from $r_{\rm circ}=1.5$ to $30.0\,R_s$. The mass of the host system is set to
$300\,m_{\rm sat}$ for these runs.

Finally, we have investigated the conditions necessary for total disruption of
an NFW halo by performing simulations of $N=10^5$ satellites on circular orbits
with orbital radius $r_{\rm circ}=1.5\,R_s$ and varying the mass of the host
system from $M_{\rm host} = 300~m_{\rm sat}$ to $M_{\rm host}=3000~m_{\rm sat}$.
The numerical and orbital parameters of these simulations are listed in
Table~\ref{tab:orbparams}.

\subsection {Mass Loss}
\label{ssec:massloss}

\begin{figure}
\epsfxsize=16.0truecm
\centerline{\epsfbox{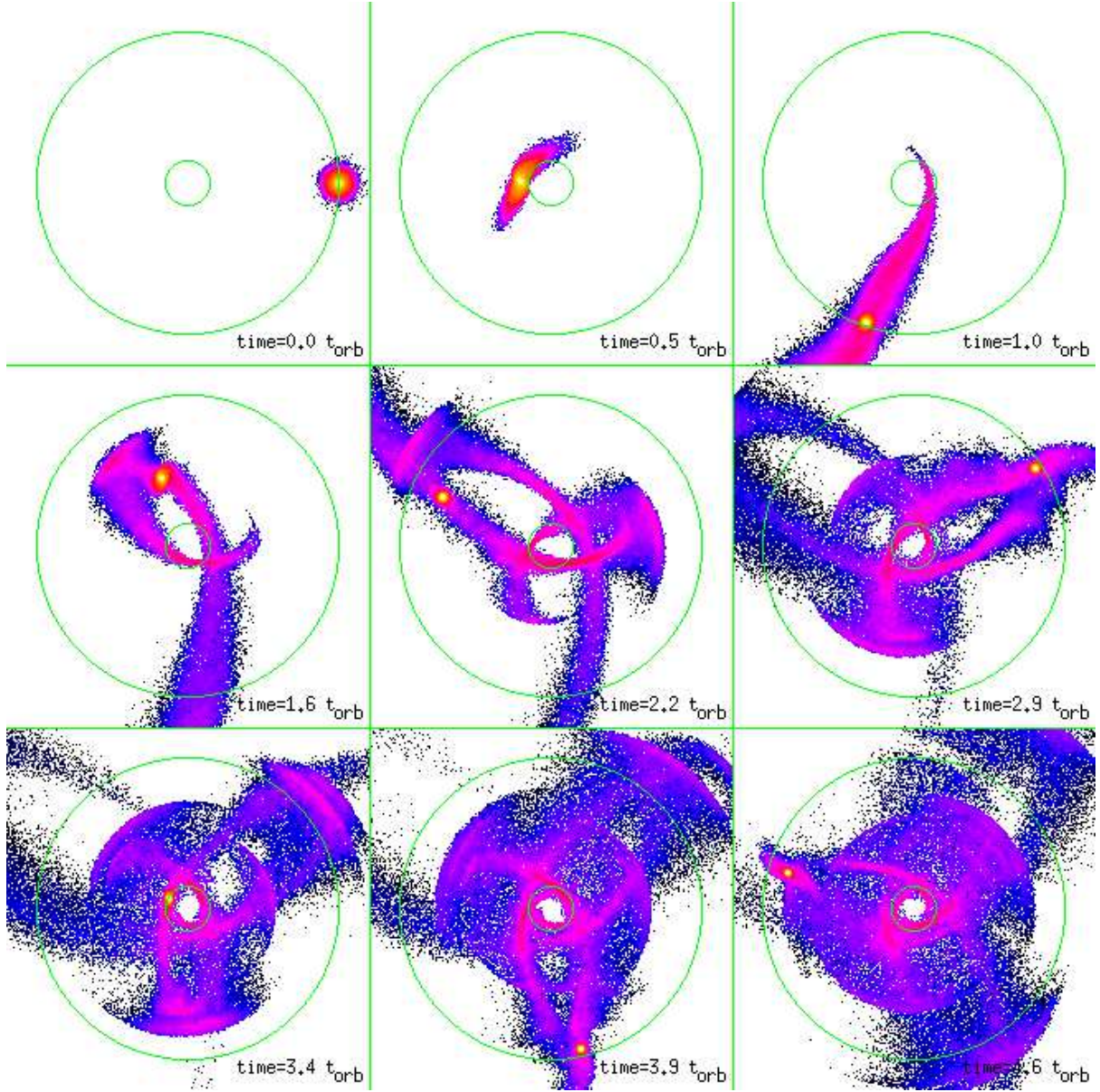}}
\figcaption{
Snapshots of satellite halo particles projected onto the orbital plane of the
satellite for the ($N,r_{\rm ap},r_{\rm per}$) = ($10^5,10.0\, R_s, 1.5\, R_s$).
The inner circle indicates the pericentric radius and the large circle denotes
the apocentric radius of the satellite.  Tidally stripped material forms
elongated tidal tails and spherical shells which extend for many scale radii
beyond the remaining bound mass of the satellite halo.\label{fig:snapshots}}
\end{figure}

Figure \ref{fig:snapshots} shows a sequence of snapshots of an $N=10^5$
satellite halo on an eccentric orbit with apocenter $r_{\rm ap}=10~R_s$, and
pericenter $r_{\rm per}=1.5~R_s$.  The effects of tidal stripping and mass loss
are clearly visible in the form of elongated tidal `tails', which extend for
many scale radii beyond the bound core of the satellite.  The satellite appears
increasingly elongated along the path of its orbit as it approaches pericenter,
and during each pericentric passage a substantial fraction of particles are
physically separated from the satellite by the tidal field of the host.

We quantify this mass loss by identifying the maximal subset of particles which
remain gravitationally self-bound.  Our procedure follows that of Tormen et
al.~(1998), and identifies iteratively all particles that have negative binding
energy relative to all others. At each iteration all escapers (i.e., particles
with positive binding energy) are removed from the system and the potential and
kinetic energies of all remaining particles (in the rest frame of the bound
subsystem) are recomputed. The iteration ends when no further escapers are
found.  Note that this process is also used by the group finder SKID to identify
gravitationally bound groups in N-body simulations (Stadel 2001).

\begin{figure}
\centerline{\epsfbox{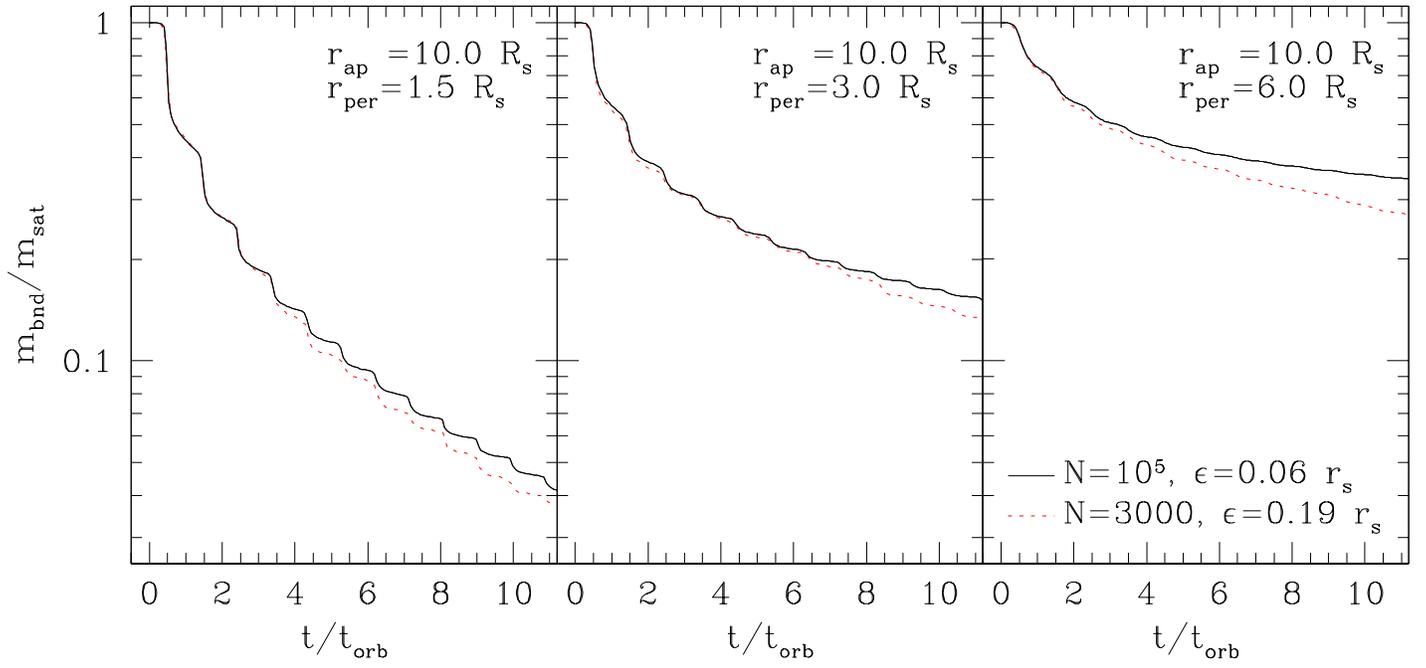}}
\figcaption{Bound mass of a satellite halo as a function of time for three
different orbits with apocentric and pericentric radii ($r_{\rm ap}$ and
$r_{\rm per}$, respectively) as shown.  Results from low and high resolution
simulations are shown together for comparison, with number of particles $N$ and
softening length $\epsilon$ as indicated.  Note the good agreement between the
two curves despite a difference of a factor of 30 in mass
resolution.\label{fig:boundmass}}
\end{figure}

The evolution of the self-bound mass, $m_{\rm bnd}$, is shown for three
different eccentric orbits in Figure \ref{fig:boundmass}. The orbital parameters
are shown in each panel. The figure shows results for models with $N=3000$
(dotted lines) and $N=10^5$ (solid lines), and confirms our conclusion in
\S~\ref{ssec:secev} that the mass loss process is rather insensitive to the mass
resolution when the number of remaining bound particles exceeds a few
hundred. Mass loss takes place in easily recognizable steps, where each event is
triggered by tides operating at pericenter. As expected, the total amount of
mass lost correlates strongly with pericentric distance; the satellite on the
$r_{\rm per}=1.5 \, R_s$ orbit is reduced to less than $5\%$ of its original
mass after $10$ orbits, while that on the $r_{\rm per}=6.0 \, R_s$ orbit retains
$35\%$ of its mass after $10$ orbits. Satellites appear to lose mass
continuously, especially for orbits with small pericentric radii, suggesting
that satellites are not in general stripped down to a finite tidal radius in a
single orbit.  Halos with the same pericentric distance may therefore be
stripped to very different degrees depending on the total number of orbits
completed.

How many orbits has a typical substructure halo completed by $z=0$? The orbital
time depends mainly on the apocentric distance (see, e.g.,
Table~\ref{tab:orbparams}), which may be assumed to scale with the time of
accretion into the host system in roughly the same way that the turnaround
radius scales with time in simple spherical infall models, $r_{ta} \propto
t^{8/9}$ (Bertschinger 1985).  For example, assuming that, for systems accreted
at $z=0$, the apocentric distance is roughly comparable to the virial radius of
the host system, and that the circular velocity of the host does not change
significantly with redshift, we estimate that halos accreted at $z=10,5,1$ have
been able to complete $83$, $32$, and $5$ orbits, respectively, in a
$\Lambda$CDM cosmogony.

Considering that it takes about $10$ orbits for a satellite to lose $\simeq
95\%$ of its original mass in an orbit with $r_{\rm per}=1.5 \, R_s$, we
conclude that it is unlikely that systems accreted after $z \sim 2.1$ have
experienced this much mass loss if placed on a comparable orbit. For orbits that
venture closer to the center than $\sim 1.5 \, R_s$ ($\sim 15$ kpc for a
galaxy-sized halo, $\sim 0.15$ Mpc for a galaxy cluster), halos must have been
accreted even later in order to retain $\gsim 5\%$ of their initial mass by the
present day.

This calculation emphasizes that substructure is continually evolving, and that
the surviving population of substructure halos may be significantly biased
towards satellites with large pericentric distances, which can survive for many
orbital times, and towards recently accreted halos, which have yet to be
experience significant mass loss. A successful model for substructure thus
require a good understanding of the statistics of accretion events, of the
orbital parameter distribution of accreting halos, as well as a reliable model
for their survival time.  We explore below whether simple models of tidal
stripping can be used to estimate accurately the effects of tides on the
evolution of the bound mass of substructure halos.

\section{Modeling Mass Loss}
\label{sec:modeling}

Over the years, a number of sophisticated theoretical prescriptions for
tidally-driven mass loss have been developed, largely within the context of the
evolution of globular clusters in external tidal fields (e.g., Gnedin et
al.~1999 and references therein; see also Taylor \& Babul 2001). However, these
techniques (e.g., Fokker-Planck models) are computationally expensive to
implement and not particularly well suited to the analysis of the evolution of
substructure in collisionless systems. Because of this, most work on the
evolution of substructure has used relatively simple prescriptions based on the
tidal approximation and on the impulse approximation to predict the effects of
tides on subhalos (Moore et al.~1996, Ghigna et al.~1998, 2000, Tormen et
al.~1998, K99a). We examine in this section the applicability of simple models
based on these approximations with the aim of deriving a simple prescription
that traces accurately, and as a function of time, the mass loss from subhalos
driven by external tides.

\subsection {Tidal Radii}
\label{sec:tidalr}

The simplest approach to mass loss modeling is based on the `tidal
approximation,' which assumes that all mass beyond a suitably defined tidal
radius is lost in a single orbit. The tidal radius, $r_t$, is usually defined as
the distance from the center of the satellite beyond which the differential
tidal forces of the host potential exceed the self-gravity of the satellite. The
precise definition of the tidal radius depends on a number of assumptions. For
example, if we assume that the gravitational potential of both the satellite and
the host are given by point masses, and that the satellite is small compared to
its distance from the center of the host, we find the definition of tidal radius
familiar from the Roche limit,
\begin {equation}
r_{\rm tR} = \left (\frac{m}{2 M}\right)^{1/3} R,
\label{eq:roche}
\end{equation}
where $m$ is the mass of the satellite, $M$ is the mass of the host, and $R$ is
their relative distance.  A more general expression may be obtained by
considering the extended mass profiles of both systems (see, e.g., Tormen et
al.~1998, K99a),
\begin {equation}
\frac{m(r_t)}{r_t^3} = \left(2 - \frac{R}{M(R)} \frac {\partial
M}{\partial R}\right) \frac{M(R)}{R^3}.
\label{eq:rtkly}
\end{equation}
A further complication is introduced by the centrifugal force experienced by the
satellite.  If this effect is considered for a satellite on a circular orbit we
find, for point masses, the Jacobi limit,
\begin {equation}
r_{\rm tJ} = \left (\frac{m}{3 M}\right)^{1/3} R.
\label{eq:jacobi}
\end{equation}
According to eqs.~\ref{eq:roche} and \ref{eq:jacobi} the mean density of the
satellite within the tidal radius is three (two) times the mean density of the
host within $R$ in the Jacobi (Roche) limit, and less than twice the mean
density of the host in the more general case of
eq.~\ref{eq:rtkly}.{\footnote{For satellites with orbital radii $R < 2.2 R_s$,
K99a suggest a smaller estimate of $r_t$ than the one given by
eq.~\ref{eq:rtkly} in order to account for tidal stripping due to resonances
between the tidal force of the host and the self-gravity of the satellite.  Of
the orbits we have simulated, this correction applies only those with $r_{\rm
per}$ or $r_{\rm circ}$ equal to $1.5\, R_s$ and is even then only a minor
($7\%$) correction.}}  Although these formulae assume circular orbits, they are
usually applied to eccentric orbits assuming that $R$ corresponds to the
pericentric distance, where tides are strongest.  Unless otherwise specified, we
will refer to the tidal limit, $r_t$, as the radius defined by
eq.~\ref{eq:rtkly}.  We prefer this definition because it takes into account the
extended mass profiles of the satellite and host, and also because it has been
used in previous analyses of substructure (Tormen et al.~1998, K99a) and
therefore it is interesting to test its accuracy.

\begin{figure}
\centerline{\epsfbox{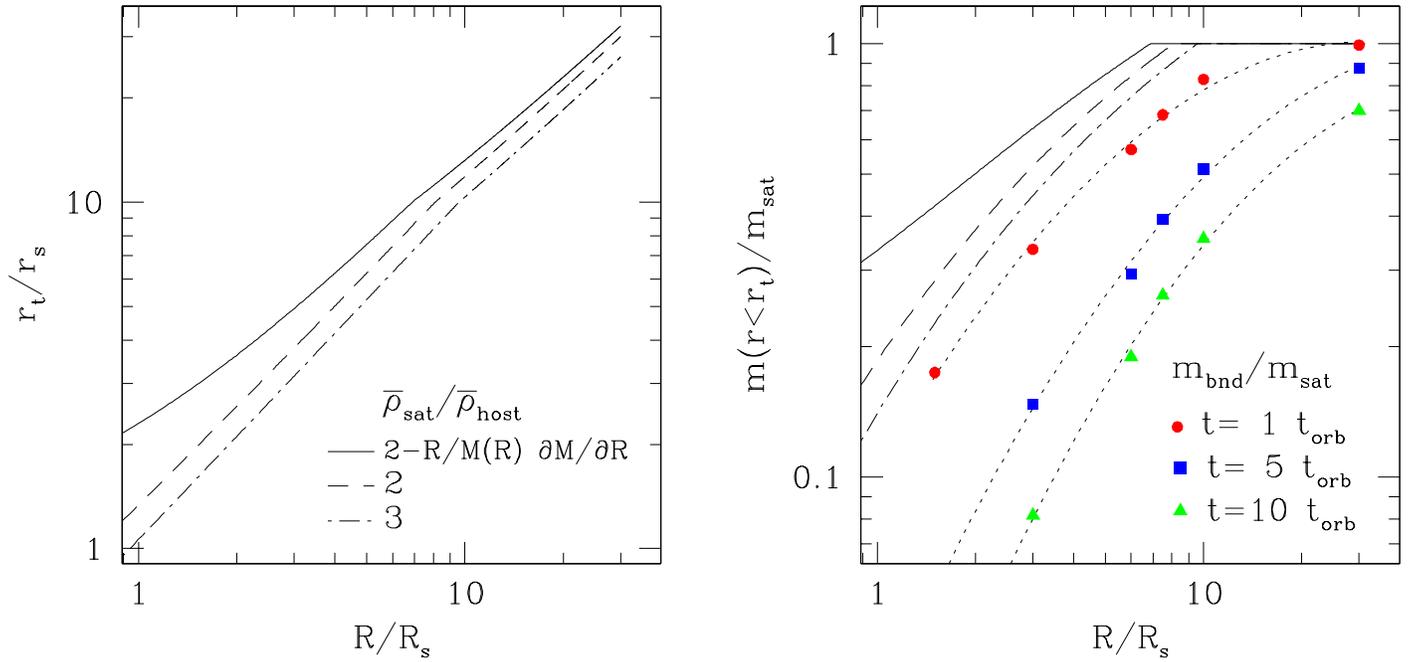}} 

\figcaption{
{\it Left}: Tidal radius as a function of orbital radius $R$ in units of the
satellite scale radius $r_s$ for $M_{\rm host}=300\, m_{\rm sat}$.  Dashed
(dot-dashed) curve shows the tidal radius corresponding to the Roche (Jacobi)
limit definition, $\overline{\rho}_{\rm sat}/\overline{\rho}_{\rm host} = 2
(3)$, where $\overline{\rho}_{\rm sat}/\overline{\rho}_{\rm host}$ is the mean
overdensity of the satellite within the tidal radius relative to that of the
host within the orbital radius.  Solid curve shows tidal radius given by
equation (\ref{eq:rtkly}). {\it Right}: Mass within tidal radius for the same
three definitions of the tidal radius.  Symbols indicate the remaining bound
mass of $3000$-particle satellites after completing $1$, $5$, and $10$ circular
orbits at orbital radii $R = 1.5, 3.0, 6.0, 7.5, 10.0$, and $30.0 \,R_s$.  Note
that even the Jacobi limit underestimates mass loss after one circular orbit by
$\sim 25\%$.
\label{fig:tidalrad}}
\end{figure}

Figure~\ref{fig:tidalrad} shows the tidal radius (left panel), as well as the
satellite mass within the tidal radius (right panel), as a function of orbital
radius for a satellite-host system with $M_{\rm host} = 300\, m_{\rm sat}$.
Three different estimates of the tidal limit are shown, the smallest being the
Jacobi limit, the largest being the radius defined by eq.~\ref{eq:rtkly}.  The
various estimates of the tidal radius vary from $r_t \simeq 1\, r_s$ to $10\,
r_s$ for orbital radii in the range $R \simeq 1 - 10\, R_s$.  Note that in all
cases the tidal radius becomes comparable to the minimum binding radius, $r_{\rm
bind}$, discussed in \S~\ref{sec:halo}, only for orbits that venture well inside
$R=R_s$, so we would expect from the simple argument presented in
\S~\ref{sec:halo} that none of the orbits with $M_{\rm host} = 300\, m_{\rm
sat}$ listed in Table~\ref{tab:orbparams} should lead to full disruption.

\subsection{Total disruption of NFW halos}
\label{ssec:totdis}

\begin{figure}
\centerline{\epsfbox{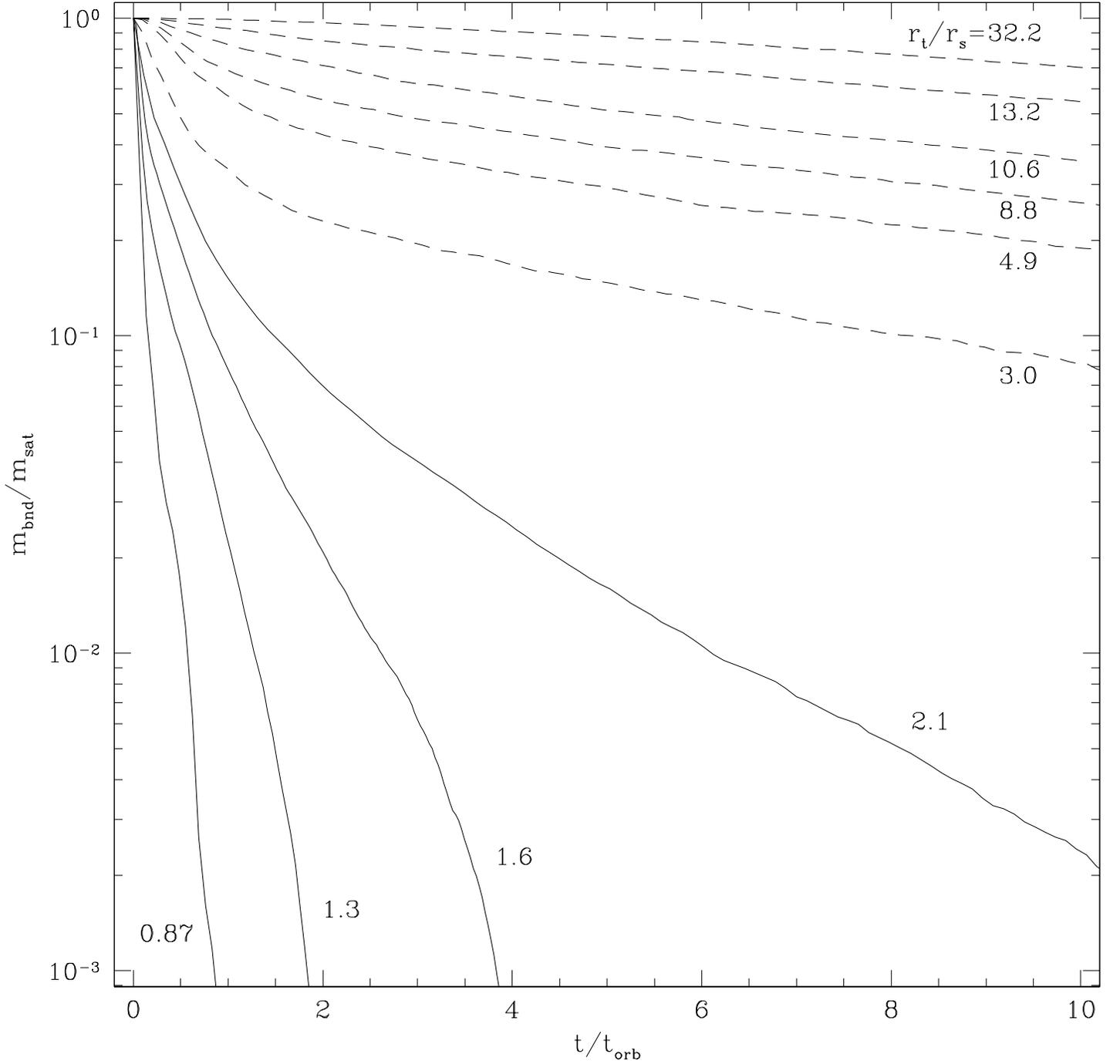}}
\figcaption{Bound mass of satellite halos on circular orbits as a function of
the number of orbits completed.  Each curve is labelled with the satellite's
tidal radius in units of the scale radius, $r_s$ (see Table
\ref{tab:orbparams}).  Solid curves indicate simulations performed with
$N=10^5$, dashed curves correspond to satellites with longer orbital periods
simulated with $3000$ particles.  Satellites with tidal radii $\lsim
2.6 \, r_{\rm
bind} \sim 2\, r_s$ are totally disrupted in a few orbits, whereas satellites
with larger tidal radii may survive self-bound for more than 10 orbits and
perhaps indefinitely.
\label{fig:simpsbmhost}}
\end{figure}

Do substructure halos subject to strong tides fully disrupt? In order to test
the conditions necessary for total disruption, we carried out a number of
$N=10^5$ simulations where the tidal radius is comparable to or smaller than the
minimum binding radius $r_{\rm bind}\approx 0.77\, r_s$. These are the circular
orbits listed in Table~\ref{tab:orbparams} with orbital radii $r_{\rm circ} =
1.5\, R_s$, and $M_{\rm host} = 300$ to $3000\, m_{\rm sat}$.  The evolution of the
bound mass of these satellites is shown in Figure~\ref{fig:simpsbmhost} by solid
curves labelled by the value of the parameter $r_t/r_s$. The dashed curves
correspond to $N=3000$ satellites with tidal radii significantly larger than
$r_{\rm bind}$.  Figure~\ref{fig:simpsbmhost} shows clearly that, for $r_t < 2\,
r_s \sim 2.6 \, r_{\rm bind}$ the satellite halo fully disrupts in just a few
orbits.  Satellites with larger tidal radii, on the other hand, lose mass at an
ever decreasing rate, and may actually survive indefinitely the effects of
tides.

This result appears to contradict the findings of Moore et al.~(2001) who claim
that ``the singular cores of substructure halos always survive complete tidal
disruption although mass loss is continuous and rapid.' '  This conclusion was
based on simulations of a Hernquist (1993) model halo on a circular orbit within
a Galactic potential, with $r_t=a$, where $a$ is the scale radius of the
Hernquist model, where the circular velocity peaks.  The minimum binding radius
of a Hernquist model occurs at $r_{\rm bind} = 0.4\, a$.  Therefore, the most
extreme orbit explored by Moore et al (2001) has $r_t \approx 2.5 \, r_{\rm
bind}$, comparable to the $r_t/r_s=2.1$ run shown in
Figure~\ref{fig:simpsbmhost}. In the latter case, a small core of $0.2\%$ of the
original mass remains self-bound even after $10$ orbits. Moore et al find that
$0.3\%$ of the original mass remains bound after about $5$ orbits, consistent
with their slightly smaller tidal radius. As shown in
Figure~\ref{fig:simpsbmhost}, smaller tidal radii lead to rapid and full
disruption. We conclude that total disruption of NFW satellites does in fact
occur, albeit only for tidal radii smaller than $\sim 2\, r_{\rm bind}$: Moore
et al concluded that halos survive indefinitely only because their simulations
did not probe tidal radii small enough to ensure disruption.

\subsection {Mass Loss and Tidal Approximation}
\label{sec:tidalml}

\begin{figure}
\centerline{\epsfbox{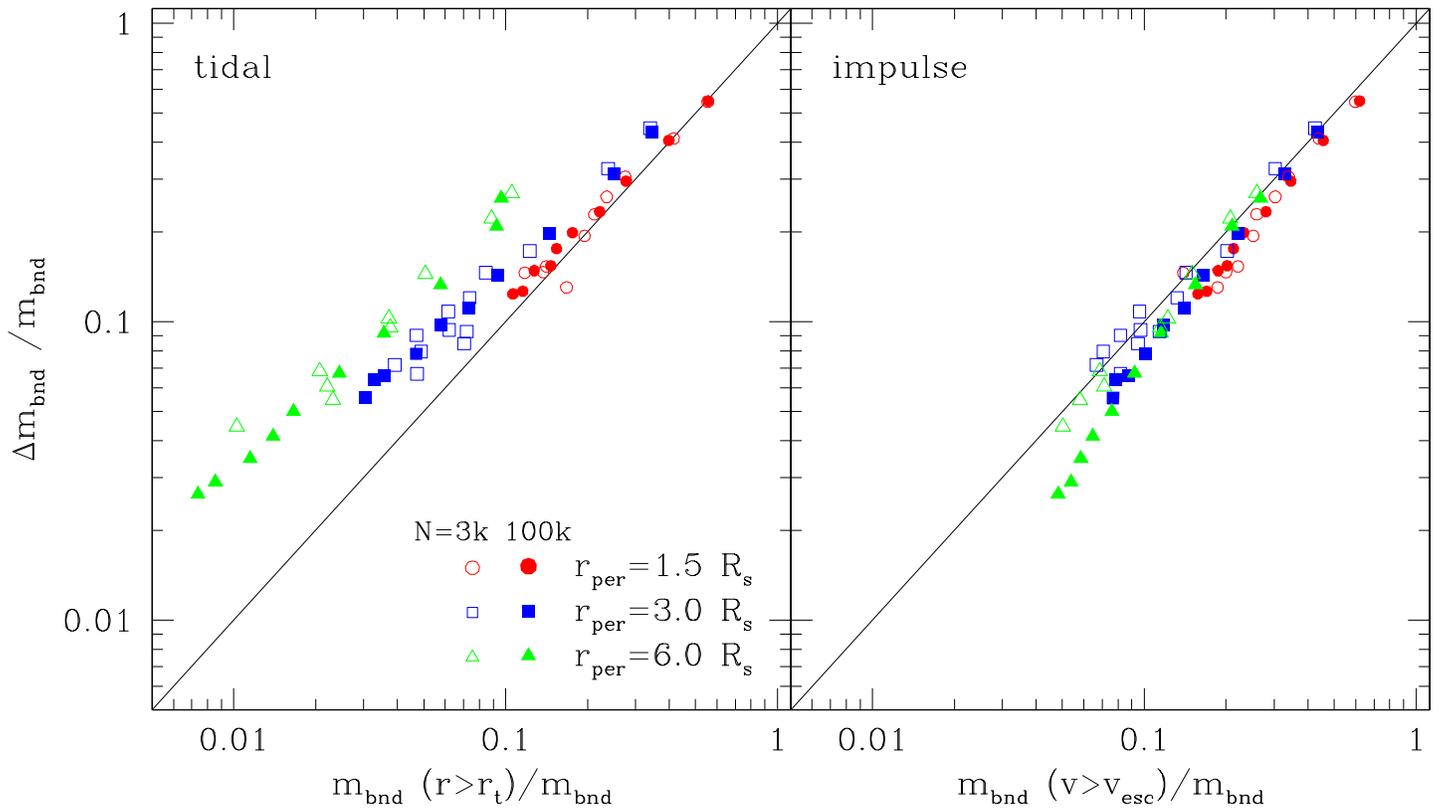}}
\figcaption{{\it Left}: Mass fraction lost 
between consecutive apocentric passages compared to predictions made using the
tidal approximation. Results for $N=3000$ and $N=10^5$ and various orbits are
shown, as labelled. Points above the solid line represent mass loss
underestimation by the tidal approximation; points below the line correspond to
overestimates.  {\it Right}: Same as left panel for impulse approximation
predictions.\label{fig:tidalimp}}
\end{figure}

How well does the tidal approximation predict mass loss during a single orbit?
This is shown in the left panel of Figure~\ref{fig:tidalimp}, where we compare
the mass fraction lost after each pericentric passage with that beyond $r_t$,
computed using the structure of the satellite measured at the preceding
apocenter. The tidal approximation is in reasonably good agreement with the
simulations in the case of strong mass loss ($\gsim 10\%$), but significantly
underestimates the effect of tides in the case of more moderate mass loss.
Comparing the open and filled symbols in Figure~\ref{fig:tidalimp} shows that
this conclusion is insensitive to the number of particles used in the
simulations.  

The tidal approximation also significantly underestimates the effects of tides
for circular orbits. This is shown in the right panel of
Figure~\ref{fig:tidalrad}, where the symbols connected by dotted lines indicate
the bound mass after completing $1$, $5$, and $10$ circular orbits at various
radii.  Even the smallest estimate of the tidal radius, i.e.\ the Jacobi limit,
underestimates the mass lost in the first orbit by $\sim 25\%$ for satellites at
$R \leq 10\,R_s$.  After $10$ orbits, satellites placed at $R=10\, R_s$ have
lost $\sim 65\%$ of their mass, although the tidal approximation would predict
that no mass loss should occur. We conclude that adopting the tidal
approximation to predict the effects of tides on substructure halos may lead to
substantial underestimation of the mass loss actually incurred by these systems.

\subsection{Impulse Approximation}
\label{sec:impulse}

An alternative approach to estimating the effects of tidal mass loss is to use
the impulse approximation to compute the changes in velocity expected for each
particle during a pericentric passage. The procedure assumes that tides operate
on a timescale short compared with the internal crossing time of the satellite,
and is therefore best suited for highly eccentric orbits, where a satellite
spends a short fraction of its orbital time near pericenter.

We follow a similar procedure to that outlined by Aguilar \& White
(1985). Freezing the configuration of a satellite at apocenter, we compute for
each particle the total change in velocity expected after one orbit if it were
to move on a trajectory similar to that followed by the center of mass of the
satellite,
\begin{equation}
{\mathbf \Delta v}_i=\int_{\rm orbit}{\mathbf a}_i(t)dt,
\end{equation}
where ${\mathbf a}_i(t)$ is the acceleration on particle $i$ due to the host
potential at each point along the orbit of the satellite. After modifying each
particle's velocity in this manner, we compute the stripped mass by applying
the same procedure described in \S~\ref{ssec:massloss}.

Unlike the tidal approximation, this method incorporates both spatial and
kinematic information about the halo particles. As seen in the right panel of
Figure~\ref{fig:tidalimp}, this extra information results in better agreement
between the mass loss predicted per orbit and the results of our numerical
experiments. The accuracy of the estimate improves as the eccentricity of the
orbit increases, as expected since the assumptions of the impulse approximation
are better justified in this case.

\subsection{Cumulative Mass Loss}
\label{sec:cumulative}

The success of the impulse approximation and, to a lesser extent, of the tidal
approximation, in predicting accurately the mass stripped in a single orbit
relies heavily on knowledge of the structure of the satellite before pericentric
approach. The results shown in Figure~\ref{fig:tidalimp} use the actual
structure of the satellite at each apocenter to predict the tidal mass loss
during the subsequent pericentric passage. Without this knowledge, the accuracy
of both analytic techniques is severely curtailed.

\begin{figure}
\centerline{\epsfbox{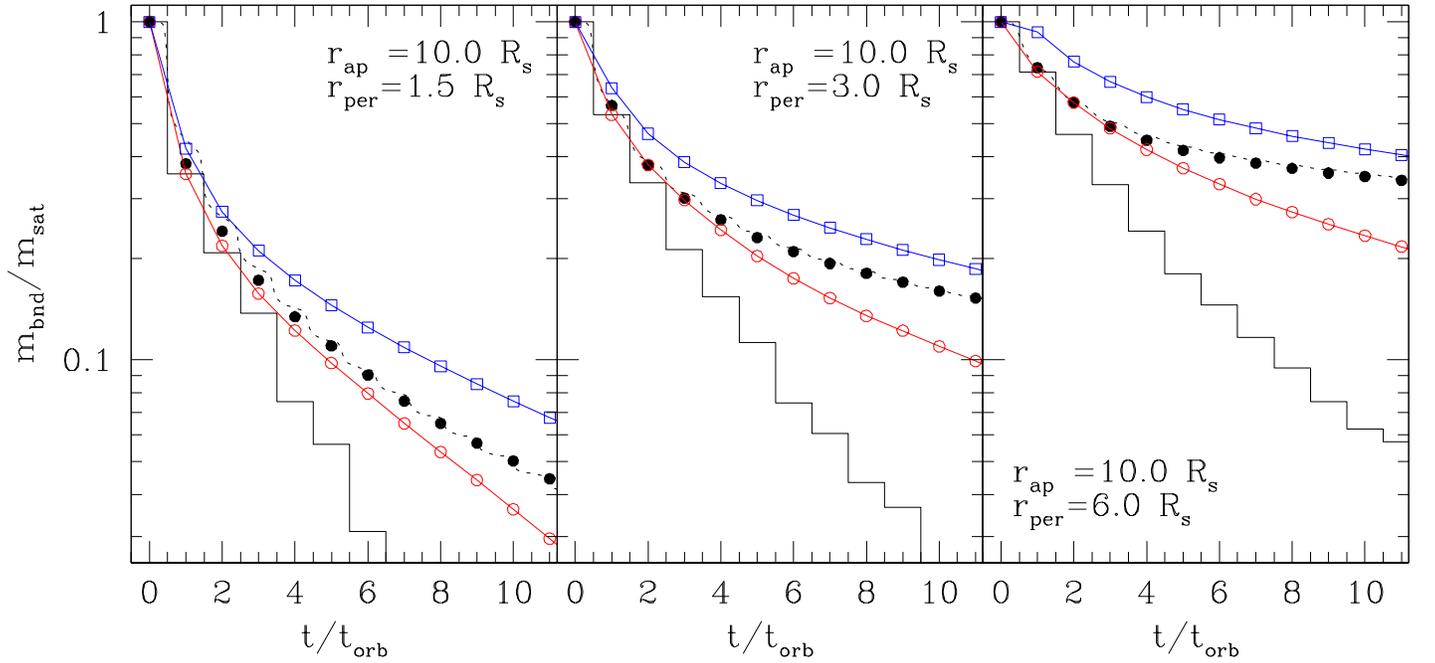}}
\figcaption{
Impulse approximation mass loss predictions compared with results for the
$N=10^5$ simulations (dotted lines) shown in Figure \ref{fig:boundmass}.
Step-like solid curves show predictions made by repeated application of the
impulse approximation.  Solid circles show independent predictions for each
orbit based on the structure of the halo at each preceding apocenter, as in
Figure \ref{fig:tidalimp}. Open symbols show predictions of tidal (squares) and
impulse (circles) approximations using the modified profile given by
eq.~\ref{eq:rhomod} to recalculate the structure of the stripped halo after each
orbit .\label{fig:mrtbothimp}}
\end{figure}

This can be seen in Figure~\ref{fig:mrtbothimp}, where we compare the results of
applying the impulse approximation with two different assumptions: (i) using
dynamical information from the actual simulation at each apocenter (filled
circles), and (ii) repeatedly applying the impulse approximation to the original
(unevolved) equilibrium configuration of the satellite (step-like solid lines
without symbols). The results of the numerical simulations are shown as dotted
lines. Clearly, unless the structural evolution after each episode of mass loss
is taken into account, the applicability of this analytic technique to the study
of the mass loss over several orbits is quite limited. The reason for this is
that the structure of the bound system evolves away gradually from the original
NFW profile as tides operate; we turn our attention to this issue next.

\section{The Internal Structure of Stripped NFW Halos}
\label{sec:stripprof}

\begin{figure}
\epsfysize=9.0truecm
%\epsfysize=0.0truecm
%\centerline{\plottwo{f10a.eps}{f10b.eps}}
%\centerline{\plottwo{f10c.eps}} 
\centerline{\epsfig{figure=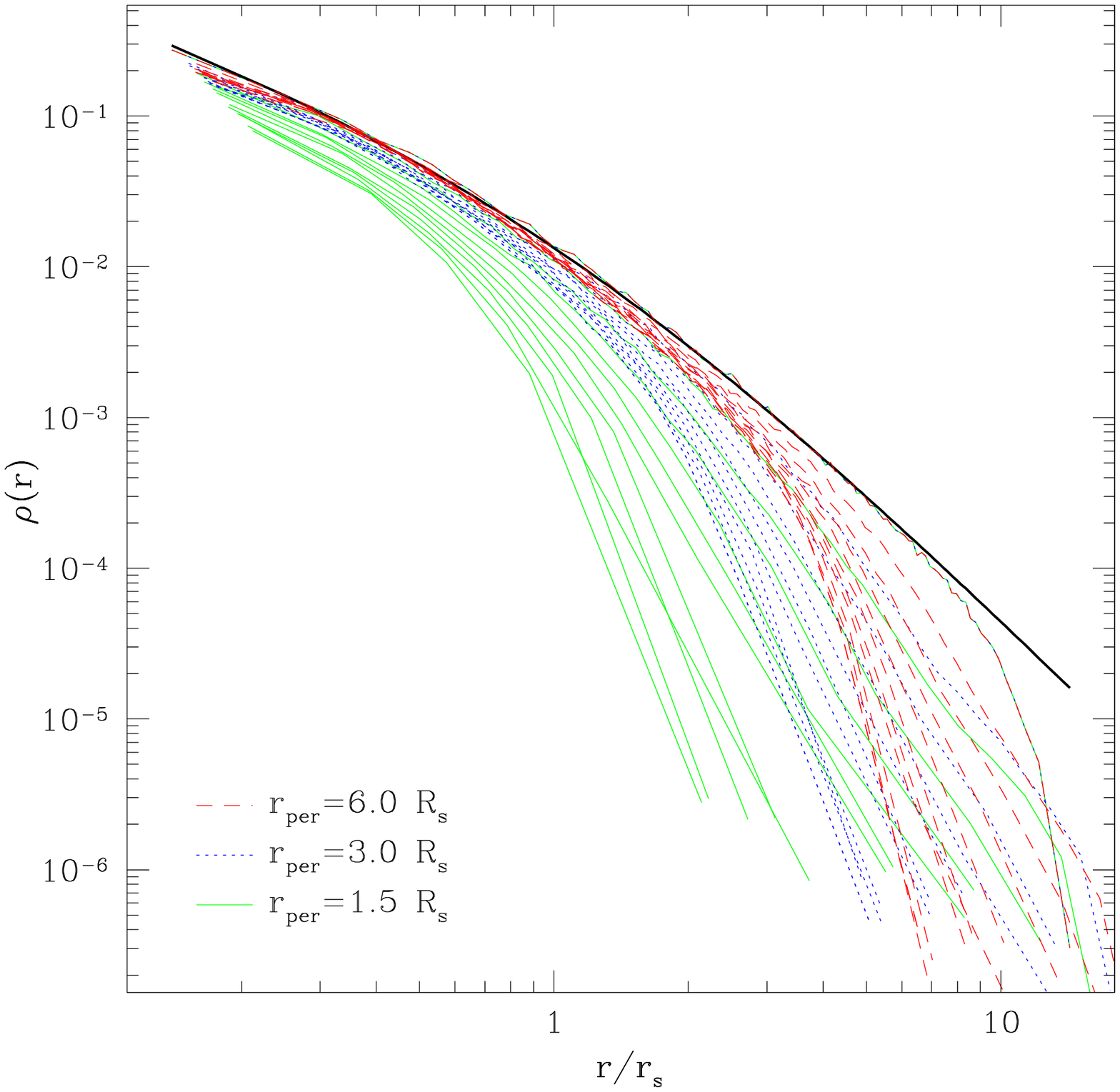,width=0.46\linewidth}\epsfig{figure=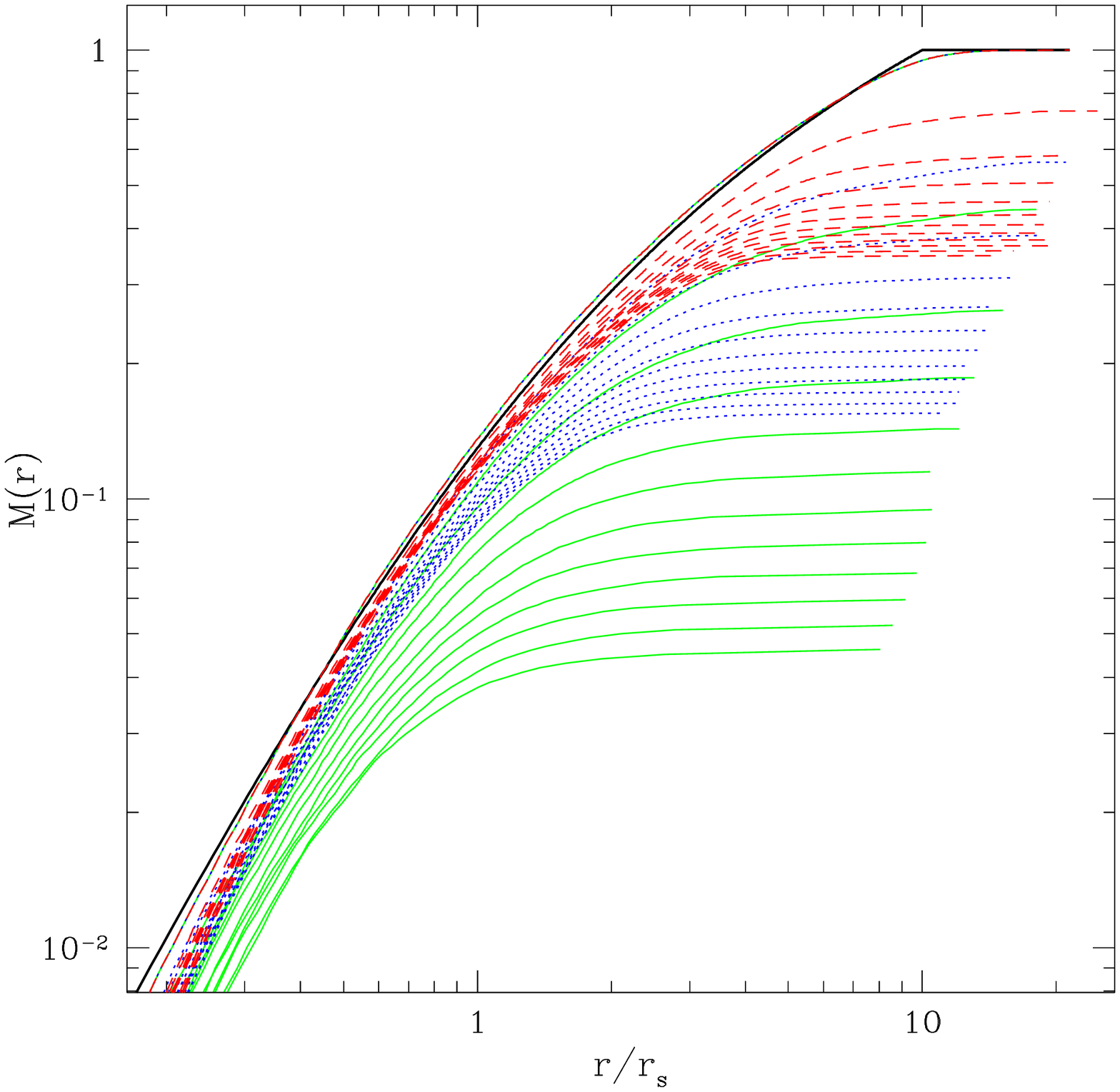,width=0.46\linewidth}}
\centerline{\epsfig{figure=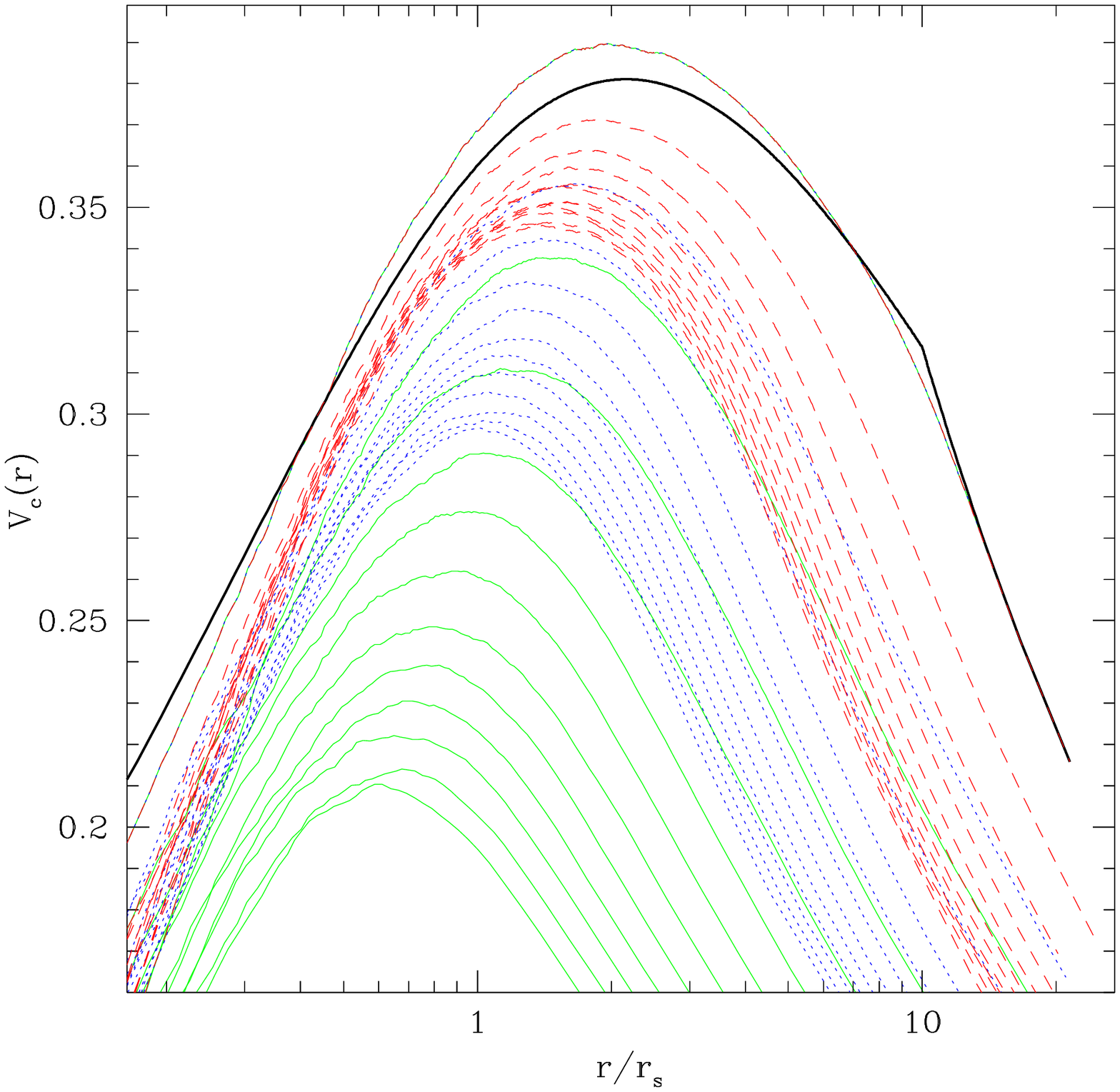,width=0.46\linewidth}} 
\figcaption{ Density, mass, and circular velocity profiles of the
bound mass of $N=10^5$ satellite halos.  Profiles are shown at
apocenter for three different orbits with $r_{\rm ap}=10.0~R_s$ and
$r_{\rm per}$ as shown.  Density is given in units of $m_{\rm
sat}/r_s^3$, mass is in units of $m_{\rm sat}$, and velocity is in
units of $\sqrt{G m_{\rm sat}/r_s}$.  Thick solid curves correspond to
the initial (unstripped) NFW profiles.\label{fig:allthree}}
\end{figure}

The structure of the bound remnant of the satellite changes steadily as the
satellite gradually loses mass through tides. This is shown in
Figure~\ref{fig:allthree}, where we plot the density, mass, and circular
velocity profiles of the bound remnant in simulations with $10^5$
particles. These profiles are shown at apocenter, when the satellite is closest
to equilibrium, and illustrate a number of interesting trends.

Firstly, as expected, much of the tidally stripped mass is lost from the outer
regions of the halo, steepening the outer density profile.  The effects of tidal
heating are, however, not restricted to the outer regions. As satellites lose
mass, the density in regions closer to the center also decreases
significantly. Note that tides do not apparently lead to the formation of a
constant-density core near the center.  These effects tend to reduce the peak
circular velocity, $V_{\rm max}$, of stripped satellites and to shift the
location of the peak, $r_{\rm max}$, inwards. The correlation between the peak
circular velocity of stripped halos and their remaining bound mass is shown in
the top panel of Figure~\ref{fig:frhor0}.  We find that $V_{\rm max}$ scales
roughly as $m_{\rm bnd}^{1/3}$ as indicated by the solid line in that figure.

The second trend to note is that the structure of a stripped halo seems to be
dictated by a single parameter: the total amount of mass lost. This is clear
from the mass- and velocity-profile panels in Figure~\ref{fig:allthree}, which
show a gradual and smooth progression in the radial distribution of mass as a
function of remaining self-bound mass.  Note that this trend is relatively
independent of the orbital parameters of the satellite, i.e., the mass profile
of a satellite that has lost almost $60\%$ of its initial mass over five orbits
is much the same as that of a satellite that has lost the same mass fraction in
a single orbit by passing much closer to the centre of the host potential.  

It is possible to describe the structure of a stripped halo through a simple
modification of the NFW profile (eq.~\ref{eq:nfw}),
\begin{equation}
\rho(r) = {f_t \over 1 + ({r/r_{te}})^3}\, \rho_{\rm NFW}(r),
\label{eq:rhomod}
\end{equation}
where $f_t$ is a dimensionless measure of the reduction in central density, and
$r_{te}$ is an `effective' tidal radius that describes the outer cutoff imposed
by tides. For $f_t=1$ and $r_{te} \gg r_{s}$, eq.~\ref{eq:rhomod} reduces to the
original NFW formula.

\begin{figure}
\centerline{\epsfbox{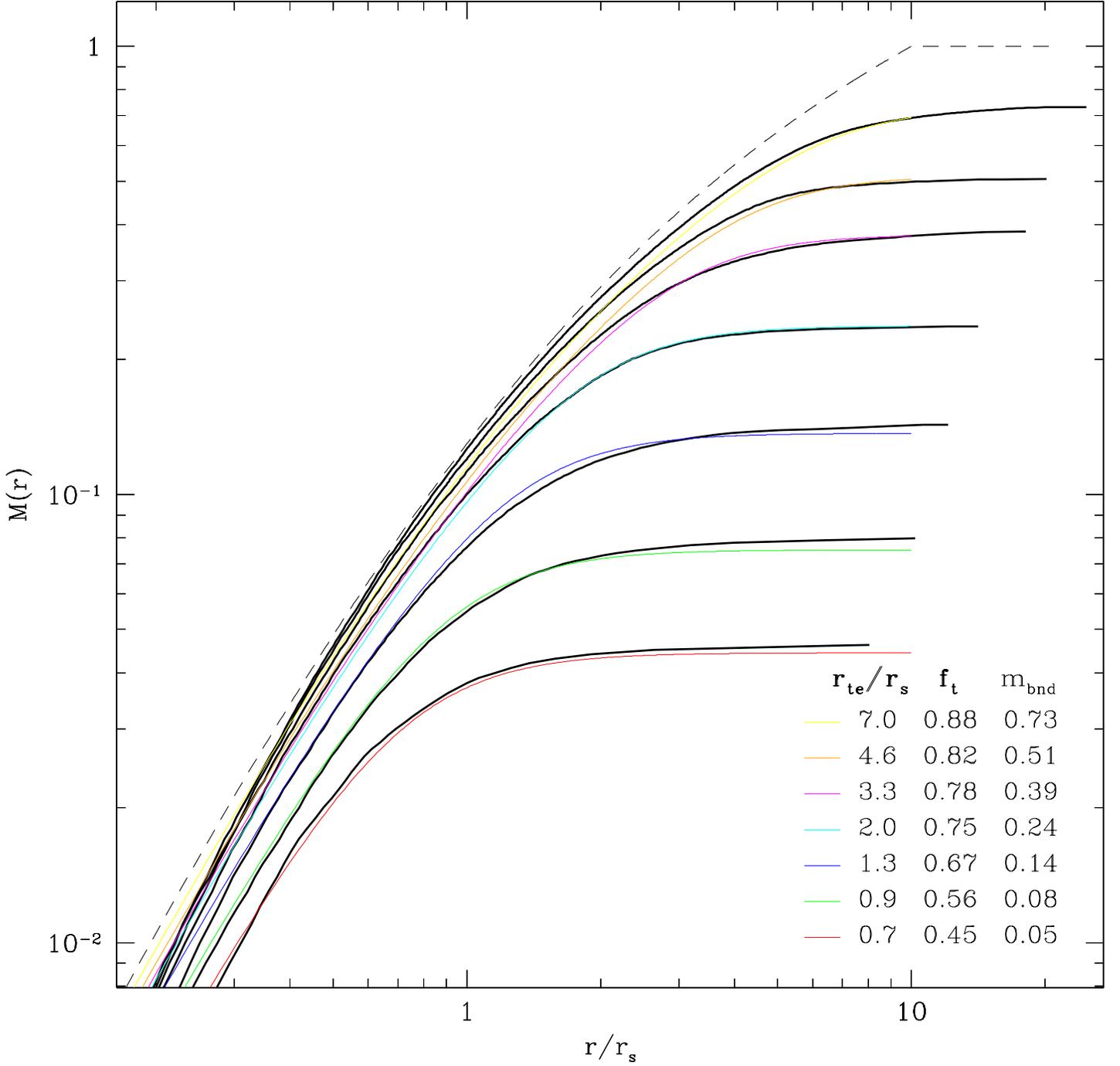}}
\figcaption{
Modified NFW profile fits to selected halo mass profiles from the middle panel
of Figure \ref{fig:allthree}.  The dashed line shows the original,
unstripped NFW mass profile.  The parameters $r_{te}$ and $f_t$
control the truncation radius and the reduction in central density of
the stripped profile, respectively.\label{fig:massfitn3}}
\end{figure}

\begin{figure}
\centerline{\epsfbox{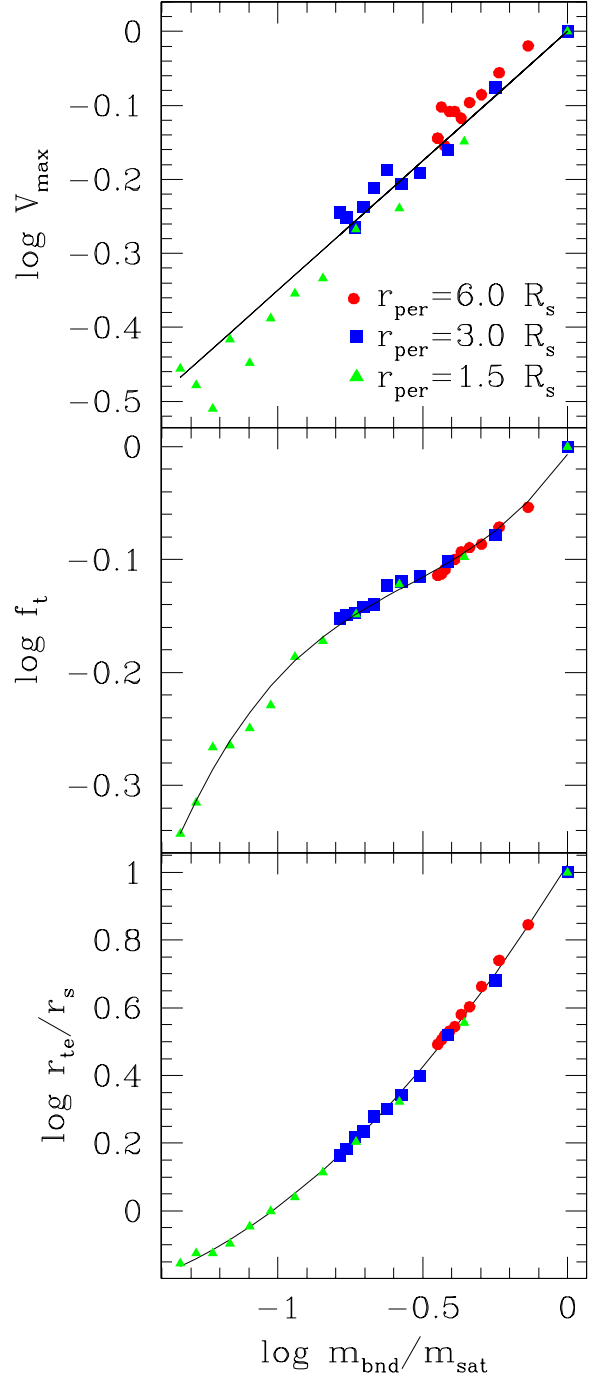}}
\figcaption{
The peak circular velocity of stripped halos, $V_{\rm max}$ (top panel) in units
of the initial peak circular velocity, as well as the parameters $f_t$ (middle
panel) and $r_{te}$ (bottom panel), shown as a function of the remaining bound
mass fraction $m_{\rm bnd}$.  Solid curves show fits to the data given by
$V_{\rm max} \propto m_{\rm bnd}^{1/3}$, and eqs.~\ref{eq:rtembnd} and \ref
{eq:ftmbnd} for top, middle and bottom panels, respectively.
\label{fig:frhor0}}
\end{figure}

Fits to various mass profiles of stripped halos using eq.~\ref{eq:rhomod} are
shown in Figure~\ref{fig:massfitn3} and are seen to describe accurately the
radial mass structure of these systems.  The parameters $f_t$ and $r_{te}$ are
not independent, and are both determined by the mass fraction of the satellite
that remains bound, $m_{\rm bnd}$, as shown in Figure~\ref{fig:frhor0}. The
effective tidal radius may be computed from the bound mass using the polynomial
fit,
\begin{equation}
\log r_{te}= 1.02 + 1.38 \, \log m_{\rm bnd} + 0.37 \, (\log m_{\rm bnd})^2,
\label{eq:rtembnd}
\end{equation}
whereas $f_t$ follows directly from $r_{te}$ by imposing the condition that the
total mass be $m_{\rm bnd}$.  In practice, $f_{t}$ is well approximated by
\begin{equation}
\begin{split}
\log f_t & = -0.007 +0.35 \, \log m_{\rm bnd} \\
& \pz \pz + 0.39 \, (\log m_{\rm bnd})^2 + 0.23 \, (\log m_{\rm bnd})^3, 
\end{split}
\label{eq:ftmbnd}
\end{equation}
as can be seen from the fits shown in Figure~\ref{fig:frhor0}.

Finally, the density-profile panel in Figure~\ref{fig:allthree} suggests that,
after a few orbits, tides tend to impose a well defined outer cutoff in the mass
distribution. This can also be seen in Figure ~\ref{fig:frhor0}, where we see
that the simulations with $r_{\rm per}=6.0\, R_s$ and $3.0\, R_s$ tend
asymptotically to a well defined minimum $r_{te}$ (and bound mass, see
Figure~\ref{fig:mrtbothimp}). The $r_{\rm per}=1.5\, R_s$ run, on the other
hand, does not participate in this trend, as it appears to approach
asymptotically total disruption.

\subsection{Cumulative Mass Loss Revisited}

We can use eq.~\ref{eq:rhomod} to predict fairly accurately the structure of the
satellite after each episode of mass loss, and use this to address analytically
the cumulative mass loss history of a satellite halo. As discussed in
\S~\ref{sec:tidalml}, this implies that we can apply either the tidal or the
impulse approximation one orbit at a time in order to predict the tidal mass
loss over several orbital periods. Our prescription for predicting cumulative
mass loss using the impulse approximation is as follows:

\begin{itemize}
\item apply the impulse approximation to an equilibrium realization of an NFW
halo and compute the mass lost after one orbit using the unbinding procedure
described in \S~\ref{ssec:massloss},

\item use the remaining bound mass to compute the modified halo profile
parameters $f_t$ and $r_{te}$ needed to calculate the mass profile of the
stripped satellite (eqs.~\ref{eq:rhomod}, \ref{eq:rtembnd} and \ref{eq:ftmbnd}),

\item construct an isotropic, equilibrium halo model with this new mass profile,

\item repeat the whole procedure using the new stripped halo model
\end{itemize}

In the case of the tidal approximation, the procedure is more efficient, since
we calculate the tidal radius using the analytic NFW and stripped halo mass
profiles instead of generating N-body realizations of halo models after each
orbit.

The success of these methods may be judged from Figure~\ref{fig:mrtbothimp},
where we show the predicted evolution of the bound mass for the three orbits
shown. The impulse approximation (open circles) reproduces the results of the
numerical simulations reasonably well, especially in the first $4-5$ orbits.
The predictions deviate from the simulation results at later times, primarily
due to discrepancies in the precise structure of the stripped halos; the change
in the shape of the profile caused by tidal heating is not perfectly captured by
the stripped halo profile of eq.~\ref{eq:rhomod}. A similar result is obtained
by applying the tidal approximation to compute the fractional mass loss during
each orbit (open squares).

In conclusion, combining the evolving structure of stripped halos
(eq.~\ref{eq:rhomod}) with simple estimates of the mass loss per orbit provides
a simple theoretical tool that may be used to guide the analysis and
interpretation of substructure studies in cosmological simulations.

\section{Substructure in the Milky Way}
\label{sec:mwsub}

The results from the previous sections indicate that substructure halos are
fully disrupted only if they venture close to the center of their host systems
and, even then, disruption takes a number of orbits to complete. It is thus not
surprising that high-resolution cosmological N-body simulations have been able
to resolve a wealth of substructure within virialized dark matter halos made up
of the partially disrupted remains of earlier accretion events (Moore, Katz \&
Lake 1996, K99a). 

What are the luminous counterparts of substructure halos?  In a galaxy cluster,
it is tempting to identify substructure halos with dynamical tracers of the
individual galaxy population, but in the case of a galaxy-sized system such
association would imply that several hundred satellites should orbit the Milky
Way. This is in stark contrast with the dozen or so known Milky Way satellites,
and implies that most subhalos must have failed to form a significant number of
stars if these models are to match observations (Kauffmann, White \& Guiderdoni
1993, M99, K99b).

\begin{figure}
\centerline{\epsfbox{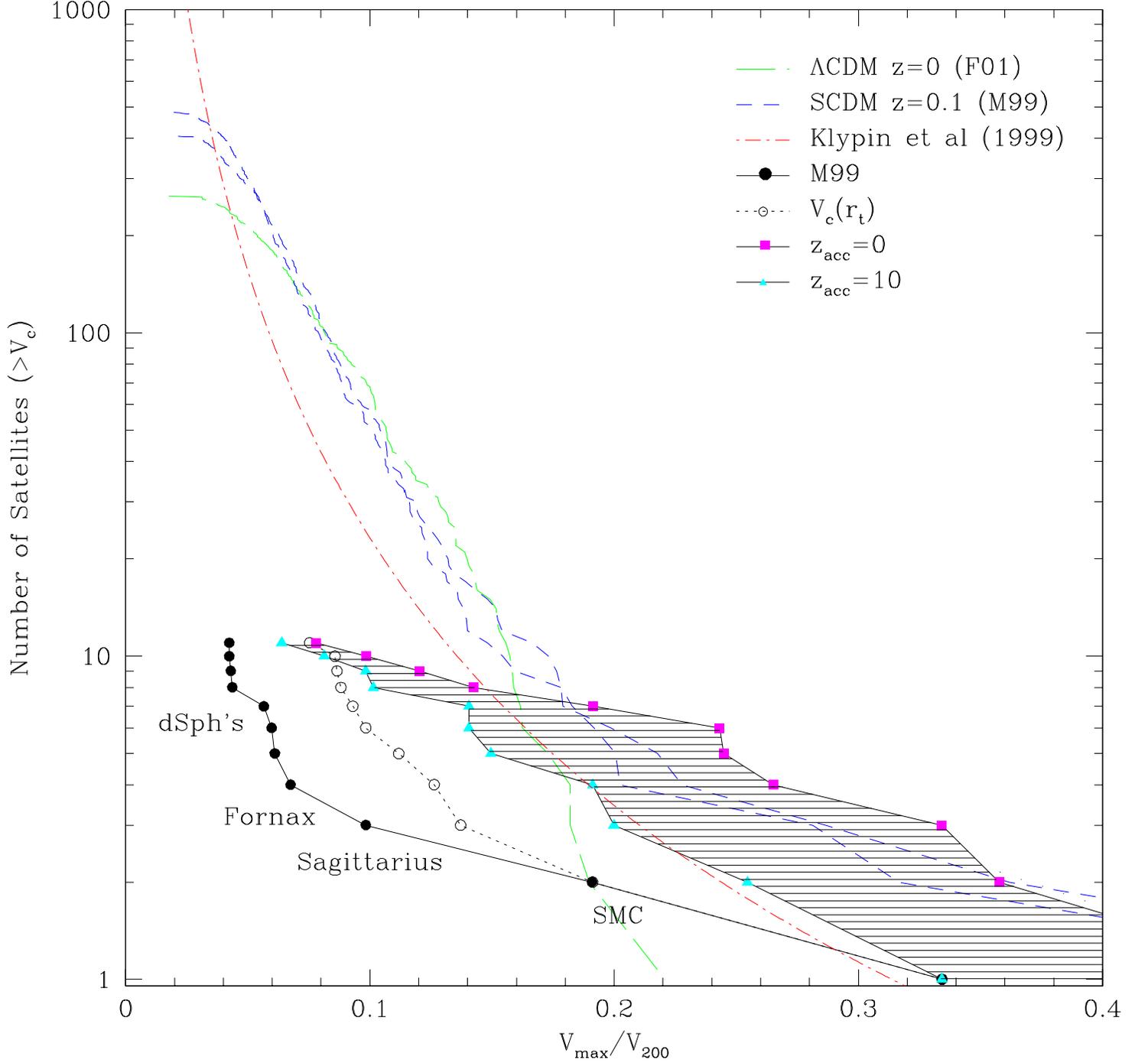}}

\figcaption{
Cumulative (peak) circular velocity function of substructure halos. Velocities
are normalized to the circular velocity of the parent halo measured at the
virial radius. The short-dashed curves correspond to two galaxy-sized halos
formed in an SCDM universe (M99). Long-dashed curves correspond to a halo formed
in the $\Lambda$CDM cosmogony (Font et al.~2001). The dot-dashed curve
correspond to results reported by K99b. Curves with symbols correspond to the
satellites observed around the Milky Way, using different assumptions to compute
the peak circular velocity of their surrounding halos. Filled circles: Milky Way
satellites assuming that stars in dwarf spheroidals are on isotropic orbits in
isothermal potentials (as in M99) and that $V_{200}=220 \kms$ for the Galaxy.
Open circles: circular velocity at the luminous cutoff ($r_{tl}$) assuming
that dwarf spheroidals are surrounded by dark halos with NFW profiles.  Shaded
region is bounded by peak circular velocity function of unstripped NFW halos
assuming two extreme accretion redshifts, $z_{\rm acc}=10$ and $0$. See text for
details.
\label{fig:vcf}}
\end{figure}

We illustrate this in Figure~\ref{fig:vcf}, where we plot, with dashed lines,
the (peak) circular velocity function of substructure halos in three
high-resolution simulations of galaxy-sized CDM halos, as compiled by Font et
al.~(2001). Scaled to the virial velocity of the host halo,
$V_{200}=V_c(r_{200})$, this function is roughly independent of the mass of the
host and of the value of the cosmological parameters, making it a robust
theoretical prediction ideal for comparison with observations of Galactic
satellites (K99b, M99, Font et al.~2001). 

Such comparison requires, however, good estimates for quantities that are not
directly accessible to observation: (i) the virial velocity, $V_{200}$, of the
Milky Way's halo, and (ii) the peak circular velocity of halos surrounding the
Galactic satellites. Conclusions from this exercise are thus sensitive to
assumptions made to infer such quantities from observables. For example, one may
assume (as in M99) that the Milky Way's halo has a virial velocity of
$V_{200}=220 \kms$, and that peak halo circular velocities in dwarf irregular
(dIrrs) satellites are identical to the rotation velocity of their gaseous
disks. In addition, peak circular velocities are inferred for dwarf spheroidals
(dSphs) by assuming that stars in these systems are on isotropic orbits in
isothermal potentials.

The result of these assumptions is shown in Figure~\ref{fig:vcf} by the filled
circles joined with a solid line. The figure includes only satellites within the
virial radius of the Milky Way, $r_{200}\approx (V_{200}/$km s$^{-1})\, h^{-1}
$kpc $\sim 338$ kpc. With these assumptions, M99 concluded that dSphs populate
halos with peak circular velocities of order $\sim 10 \kms$ ($V_{\rm max}\sim
0.05\, V_{200}$). According to Figure~\ref{fig:vcf}, one expects several hundred
substructure halos of comparable velocity, implying that fewer than one in ten
of such halos are inhabited by luminous dSphs.

What makes the very few systems that host luminous dSphs distinct from the rest?
This question has prompted suggestions that dwarf spheroidals formed only in
subhalos that collapsed before the universe was fully reionized, since
late-collapsing systems would experience difficulty retaining and cooling their
baryonic component after reionization (Bullock, Kravtsov \& Weinberg 2000,
Somerville 2001, Benson et al.~2001).

White (2000), on the other hand, has questioned this scenario on the grounds
that the isothermal assumption used by M99 to derive peak circular velocities
for dSphs is not supported by the results of direct numerical simulations. As
discussed in \S~\ref{sec:halo}, numerical simulations indicate that halo
circular velocities decrease systematically towards the center, implying that if
stars populate the innermost regions of subhalos, then the isothermal assumption
may substantially underestimate the peak circular velocities of their
surrounding halos. Assuming that dSphs are surrounded by NFW halos, White (2000)
estimated that dSphs may plausibly inhabit potential wells with circular
velocities up to a factor of $3$ times larger than inferred under the isothermal
assumption. A correction of this magnitude would reconcile, at the high mass
end, the Milky Way satellite velocity function with the subhalo velocity
function (shown with dashed lines in Figure~\ref{fig:vcf}). We emphasize,
however, that the magnitude of the correction depends sensitively on the inner
structure of subhalos, which is poorly resolved even in the best cosmological
simulations available at present.

Resolving this controversy is important for a couple of reasons. Firstly, if
dSphs indeed inhabit halos with peak circular velocities as high as $30 \kms$
($\sim 15\%$ of $V_{200}$), then essentially all such halos would host a dSph
and the scarcity of dSphs would just reflect the relative scarcity of massive
substructure halos. This would suggest that reionization has not played a major
role in hindering the formation of dwarf spheroidal galaxies. The second reason
is that if dSphs are indeed surrounded by halos that massive, it would be easier
to understand how they have managed to retain their gaseous components so as to
undergo several distinct episodes of star formation (Mateo 1998).  We use below
our results for the structure of stripped substructure halos in order to place
constraints on the peak circular velocity of dark halos surrounding dwarf
spheroidals.

\subsection{The Dark Halos of Dwarf Spheroidals}
\label{ssec:dsph}

If the shape of the dark halo mass profile is specified, it is possible to use
the velocity dispersion and spatial distribution of the stellar component to
determine the structural parameters of halos surrounding dSphs. Assuming
spherical symmetry and that the mass profile of substructure halos can be
approximated by eq.~\ref{eq:rhomod}, we may use Jeans' equations to estimate the
halo parameters. The velocity dispersion, $\sigma_*$, of stars embedded in a
system with circular velocity profile $V_c(r)$ is given by
\begin{equation}
\sigma_*^2={1\over 3 M_*} \int V_c^2(r) {dM_* \over dr} dr,
\label{eq:sigmastar}
\end{equation}
where $M_*(r)$ is the stellar mass profile. As is customary, we use theoretical
King models (King 1966) to describe the dSph luminosity profiles; the parameters
used for each system are listed in Table~\ref{table:dsph}, together with the
appropriate references. The analysis assumes a stellar mass-to-light ratio of
$2$ (solar units) in the $V$-band, although we note that our conclusions are
rather insensitive to plausible choices of this parameter.

The dark halo contribution to the circular velocity profile in
eq.~\ref{eq:sigmastar} is fully specified once three quantities are defined: the
physical parameters of unstripped NFW halos, $\rho_s$ and $r_s$ (see
eq.~\ref{eq:nfw}), as well as the total fraction of mass lost (if any). The NFW
parameters depend on cosmogony, and have been selected assuming a $\Lambda$CDM
cosmogony ($\Omega_0=0.3$, $\Omega_{\Lambda}=0.7$, $h=0.65$, $\sigma_8=0.9$),
following the prescription of Eke, Navarro \& Steinmetz (2001).

\subsubsection{Carina}
\label{sssec:carina}

\begin{figure}
\centerline{\epsfbox{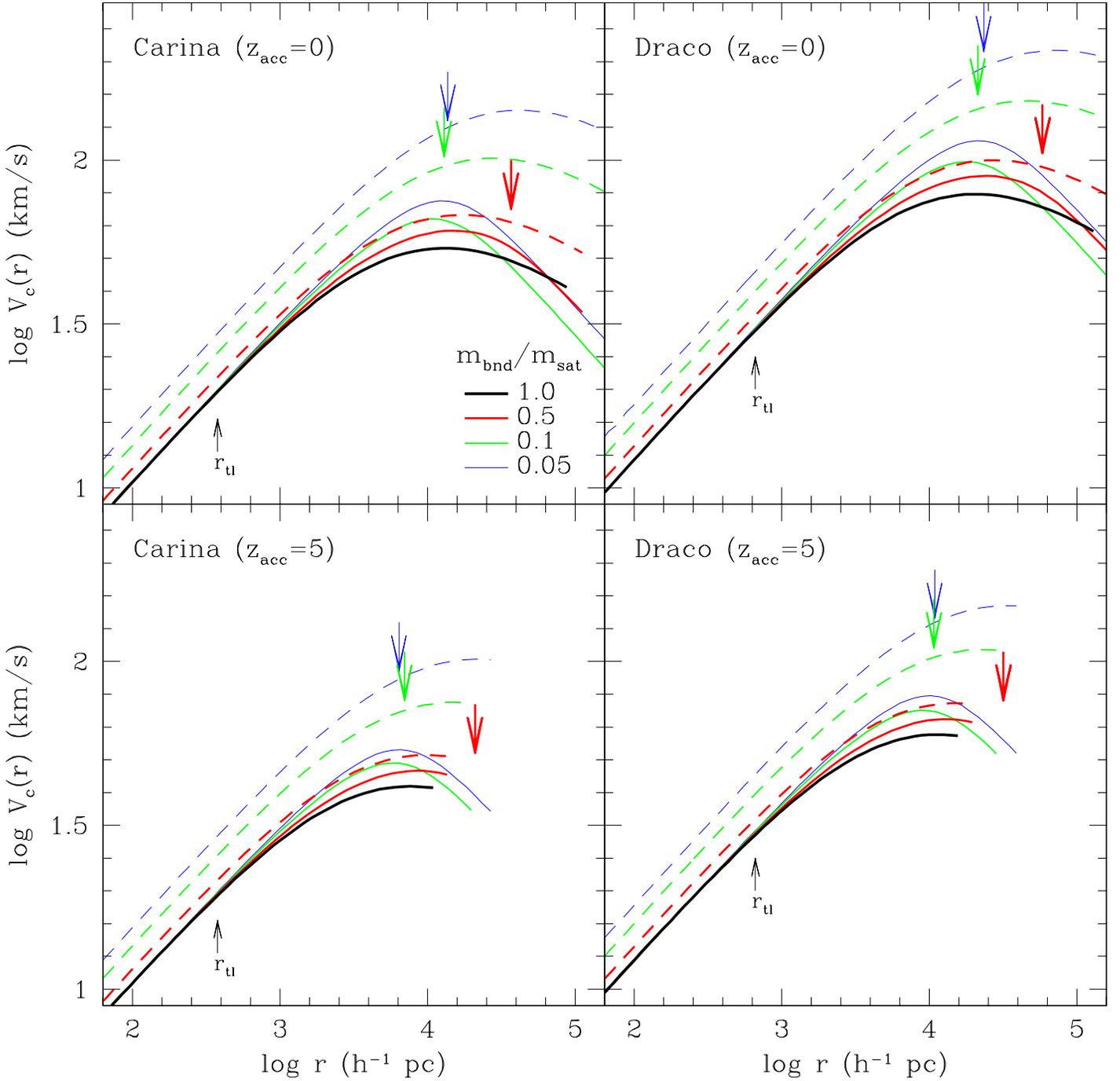}}
\figcaption{Circular velocity profiles of unstripped and stripped halos chosen
to match the surface brightness profile and the stellar velocity dispersion of
Carina (left panels) and Draco (right panels).  Dashed curves show the circular
velocity profiles of the corresponding halos before stripping.  Upward vertical
arrows indicate the luminous cutoff radius $r_{tl}$.  The luminous cutoff radius
shown for Draco is that of Odenkirchen et al.~(2001), who find $r_{tl} = 1020$
pc.  Downward vertical arrows indicate the tidal radius $r_{te}$ corresponding
to each of the stripped halos.  Top and bottom panels show halo profiles
adopting Eke, Navarro \& Steinmetz (2001) NFW parameters corresponding to halos
identified at $z=0$ and $5$, respectively.
\label{fig:carina}}
\end{figure}

The solid curves in the top left panel of Figure~\ref{fig:carina} illustrate the
circular velocity profile of (stripped) halos chosen to match the observed
stellar velocity dispersion ($\sigma_* \approx 6.8 \kms$) of the Carina dwarf
spheroidal. Different curves correspond to various assumptions about the
fraction of mass that has been lost to stripping, ranging from $95\%$ to a case
where no mass loss has occurred (thick solid line).  The dashed curves in the
same figure show the circular velocity profiles of the corresponding halos
before stripping.  The NFW concentration parameter ($c \equiv r_{200}/r_s$) of
the halos before stripping varies from $c \sim 14$ in the case of no mass loss
to $c \sim 12$ in the case of $95\%$ mass loss.  An upward vertical arrow
labelled $r_{tl}$ indicates the location of the `tidal' cutoff in the luminous
profile, obtained from the literature (see Table~\ref{table:dsph}). Downward
vertical arrows indicate the location of the tidal radii, $r_{te}$,
corresponding to each stripped halo.

Figure~\ref{fig:carina} shows that the stellar velocity dispersion constrains
tightly the circular velocity within the `tidal' cutoff of the stellar
component, $V_c(r_{tl})$, regardless of the degree of stripping. Indeed, all
circular velocity curves that match this constraint (solid lines in
Figure~\ref{fig:carina}) agree closely at $r_{tl}$, regardless of the stripped
mass fraction, and have roughly the same peak circular velocity. This implies
that the peak velocity of Carina's halo is well defined within this model, and
is unlikely to be less than $54 \kms$, well in excess of its stellar velocity
dispersion of $6.8 \kms$.

An important consequence of this conclusion is that the true tidal radius of
Carina's halo should greatly exceed the cutoff in the surface brightness profile
that is usually interpreted as a tidal limit. Even assuming that $95\%$ of the
original halo mass has been stripped the true tidal radius is still $\sim 20$
times larger than derived from King model fits to the luminosity profile. This
suggests that the stars beyond the luminous cutoff detected, for example, by
Majewski et al.~(2000), may not actually represent a true population of unbound,
extra-tidal stars but rather correspond to a radially extended component {\it
bound} to Carina. Confirmation of the extra-tidal nature of such stars through
independent means, such as detecting tidal tails, or apparent `rotation' in
Carina's extended envelope (see Majewski et al.~2000 for details) would thus
provide a strong argument against this conclusion.

We note that Mayer et al (2001) have recently reached a similar conclusion,
noting that if Carina's luminous cutoff is truly tidal in origin this would be
very difficult to reconcile with the massive dark halos expected to surround
dwarf spheroidals. Mayer et al's conclusion relies on their identification of
dSphs as tidally-stirred dIrrs; our results, on the other hand, are independent
of the formation mechanism for dSphs and rely solely on the assumption that
stars are in dynamical equilibrium within a dominant dark matter halo, an
assumption well founded on current interpretation of the observational
evidence. Our conclusion is also supported by the work of Stoehr et al (2002),
who argue that the internal structure and kinematics of the Milky Way satellites
are in excellent agreement with substructure found in high-resolution CDM
simulations provided that the satellites inhabit the most massive substructure
halos.

\subsubsection{Draco}
\label{sssec:draco}

The above conclusion may also apply to other dSphs, where apparent radial
cutoffs in the outer surface brightness profiles could actually be of little
dynamical significance and may not necessarily correspond to actual tidal
features. One especially interesting case is Draco. Using the core and tidal
radii derived by Irwin \& Hatzidimitriou (1995, hereafter IH95) from
photographic material ($r_{\rm core}=158$ pc, $r_{tl}= 498$ pc), our analysis
requires a very large halo peak velocity, $V_{\rm max} \sim 130 \kms$, to
accommodate the observed stellar velocity dispersion ($\sigma_* \sim 9.5 \kms$,
Mateo 1998).  This uncomfortably large velocity, almost comparable to that
typically associated with $L_*$ galaxies, is difficult to understand given the
low luminosity of a dwarf system such as Draco.

One possibility is that gas dynamics and star formation have significantly
altered the mass profile of the halo so that it is no longer well-described by
an NFW model.  Assuming an isothermal halo model with a flat circular velocity
profile as in M99, for example, yields $V_{\rm max} = \sqrt{2} \sigma_* = 13.4
\kms$ for Draco.  Given the assumption of NFW halo models, however, the main
reason for the large peak velocity is the rather small tidal radius derived for
Draco by IH95; this implies a very large dark matter density within $r_{tl}$ to
account for the observed velocity dispersion, which in turn requires a very
massive halo. (Within a given radius, more massive NFW halos are also denser, as
shown by the dashed lines in Figure~\ref{fig:carina}; see also NFW and Eke,
Navarro \& Steinmetz 2001.)

A similar conclusion was reached by Burkert (1997), who argued that the mean
density within $r_{tl}$ in Draco is much higher than the mean Galactic density
interior to Draco's distance to the center of the Galaxy and that, therefore,
the cutoff radius derived by IH95 cannot represent a true limitation imposed by
tides. We emphasize, however, that our conclusion is independent of Burkert's
analysis, and is based solely on our results for the structure of stripped NFW
halos, as discussed above.

Our results thus support the revision to Draco's luminosity profile derived by
Odenkirchen et al.~(2001) on the basis of data from the Sloan Digital Sky Survey
(SDSS). These authors fail to find a well defined tidal feature in Draco's
surface brightness profile and conclude, by fitting empirical King profiles
(King 1962) that, if one exists, it must lie beyond $40.1$ arcmin, more than
$40\%$ farther from the center than IH95's original determination of $28.3$
arcmin (along the major axis). Fitting theoretical King models (King 1966)
results in an even larger limiting radius, $r_{tl} \gsim 49.4$ arcmin.  

Repeating our analysis with the SDSS theoretical King model parameters ($r_{\rm
core}=179$ pc, $r_{tl}=1020$ pc and assuming the distance to Draco as
Odenkirchen et al.~(2001) ($d = 71 \pm 7$ kpc), we find a peak circular velocity
of order $79 \kms$ for Draco, more in line with what might be expected from the
very low luminosity of this system (top right panel of Figure~\ref{fig:carina}).
The concentration parameter of the Draco model halos varies from $c \sim 14$ in
the case of no mass loss to $c \sim 11$ in the case of $95\%$ mass loss.
Interestingly, we also find good agreement with the recent work of Kleyna et
al.~(2001), who find a mean mass-to-light ratio within $\simeq 30$ arcmin of
$(330 \pm 125) M_\odot/L_\odot$.  Within the same radius, our model gives $M/L =
346 \pm 128 M_\odot/L_\odot$.

\subsection{Dependence on Accretion Redshift}
\label{ssec:zdep}

As the discussion above illustrates, the luminosity profile and velocity
dispersion of dSphs constrain tightly the circular velocity within the luminous
boundary of the system, $V_c(r_{tl})$. For unstripped NFW halos identified at a
given redshift, this constraint is sufficient to determine the virial mass of
the halo: our analysis in \S~\ref{ssec:dsph} assumes NFW parameters appropriate
for halos identified at $z=0$ and shows that the $V_c(r_{tl})$ constraint can only
be matched if fairly massive halos surround dSphs. Since tides remove mass from
all radii, the mass of unstripped NFW halos that match $V_c(r_{tl})$ provide a
lower limit to the mass of halos where dSphs formed.

It is possible, however, that the characteristic densities and scale radii of
dSph halos are better described by the parameters of NFW halos identified at
higher redshift. This is because substructure halos are expected to evolve
differently from the relatively isolated systems dealt with by the numerical
work on which our estimates of $\rho_s$ and $r_s$ are based (see, e.g., NFW,
Eke, Navarro \& Steinmetz 2001). In particular, subhalos effectively stop
accreting mass after being incorporated into a more massive system, implying
that their structural parameters may very well reflect the NFW concentrations
prevalent at the redshift of accretion rather than at present.

How does this affect the peak circular velocity derived for dSph halos? Dark
halos are denser at higher redshift, so it is in principle possible for lower
mass halos to match the $V_c(r_t)$ constraint. This is shown for Carina in the
bottom left panel of Figure~\ref{fig:carina}, which is identical to the top left
panel but adopting NFW parameters appropriate for halos identified at $z=5$
rather than at $z=0$.  The concentration parameter is $c \sim 3.0$ both cases.
The effect is noticeable, but weak; the minimum peak velocity compatible with
the $V_c(r_{tl})$ constraint for Carina is $53.8 \kms$ for $z=0$, and $41.6
\kms$ for $z=5$.  The bottom right panel of Figure~\ref{fig:carina} shows the
same effect for Draco.  The concentration parameter here varies from $2.9$ in
the case of no mass loss to $2.5$ in the case of $95\%$ mass loss.  The minimum
peak velocity is $79 \kms$ for $z=0$, and $60 \kms$ for $z=5$.

Thus, a firm lower limit on the peak velocity of dSph halos requires an upper
limit on the accretion redshift of dSphs into the Milky Way's halo.  The
accretion redshift itself is highly uncertain, but it is possible to derive a
rough upper limit by considering the present-day Galactocentric distance of a
dSph. As discussed in \S~\ref{ssec:massloss}, the spherical infall model
suggests that satellites accreted early should orbit closer to the Galaxy than
late accreting systems. In other words, the present-day Galactocentric distance
of a satellite provides a rough upper limit to the virial radius of the Galaxy
at the time of accretion. 

Given the one-to-one correspondence between virial radius and accretion time
implied by the spherical secondary infall model (\S~\ref{ssec:massloss}), one
can use this assumption to derive accretion redshifts for all dSphs.  For
example, a satellite presently at the virial radius would be assigned an
accretion redshift of $z_{acc}=0$; the same procedure assigns $z_{acc}\sim 3.3$
to Ursa Minor, the closest dSph at $68$ kpc, and $z_{acc}\sim 0.6$ to Leo I, the
farthest in our sample at $254$ kpc. These are best regarded as {\it upper
limits} to the true accretion redshift: the true apocenter may lie well beyond
the present-day Galactocentric distance. Likewise, the semimajor axis of the
satellite's orbit may have been eroded by dynamical friction. Both effects would
reduce the actual accretion redshift relative to our estimate.

\subsection{The Circular Velocity Function of dSph Halos}

We list in Table~\ref{table:dsph} the {\it upper limits} to the accretion
redshift ($z_{acc}$) derived for each dSphs in our sample using the procedure
described in the previous subsection. These can in turn be used to derive firm
{\it lower limits} to the peak circular velocity of their surrounding halos,
listed as $V_{\rm max}(z_{acc})$ in Table~\ref{table:dsph}. To be conservative,
we adopt $z_{acc}$=10 for all dSphs (well in excess of the upper limits shown in
Table~\ref{table:dsph}) and derive the circular velocity function of Milky Way
satellites. This is shown in Figure~\ref{fig:vcf} as the leftmost boundary of
the shaded region; the rightmost boundary corresponds to assuming $z_{acc}=0$
for all satellites. 

Figure~\ref{fig:vcf} thus confirms the suggestion of White (2000) that in all
likelihood the circular velocity of dSph halos peaks at values that greatly
exceeds the circular speed at the luminous boundary $V_c(r_{tl})$, shown as open
circles connected by a dotted line. This implies that their dark halos must
extend well beyond $r_{tl}$. The shaded region is intended to illustrate the
uncertainty in this function associated solely with the accretion redshift
dependence of our $V_{\rm max}$ estimates (true uncertainties could be
substantially larger). Since these estimates neglect any possible stripping,
they truly represent lower limits to the actual peak circular velocity of dSph
halos.

Even under such conservative assumptions, we can see from Figure~\ref{fig:vcf}
that, within the uncertainties, there appears to be no major discrepancy between
the number of {\it massive} satellites expected in the CDM scenario and the
known satellite companions of the Milky Way. Only satellites with peak circular
velocities $\lsim 35$ km/s ($\lsim 0.16 \, V_{200}$) seem to lack readily
detectable luminous components. The scarcity of satellites around the Milky Way
reflects the small number of substructure halos with circular velocities
exceeding $\sim 35$ km/s, and a simple scenario where feedback prevents galaxies
from forming in systems below a critical circular velocity seems consistent with
observation. Our results thus indicate that it may be possible to explain the
abundance of dSphs without invoking a substantial role for the reionization of
the universe.

Even under such conservative assumptions, we can see from Figure~\ref{fig:vcf}
that, within the uncertainties, there appears to be no major discrepancy between
the number of {\it massive} satellites expected in the CDM scenario and the
known satellite companions of the Milky Way. Only satellites with peak circular
velocities $\lsim 35 \kms$ ($\lsim 0.16 \, V_{200}$) seem to lack readily
detectable luminous components. The scarcity of satellites around the Milky Way
thus reflects the small number of substructure halos with circular velocities
exceeding $\sim 35 \kms$.

\section {Summary}
\label{sec:concs}

We have used N-body simulations to investigate the structural evolution of
substructure halos orbiting within a massive host system modeled as a static
potential. We assume that the equilibrium structure of unstripped halos can be
approximated by the density profile proposed by NFW and that dynamical friction,
halo-halo encounters, and departures from spherical symmetry play a minor role
in the evolution.

Our main conclusions may be summarized as follows.

\begin{enumerate}

\item The structure of N-body realizations of equilibrium NFW halo models evolve
significantly as a result of discreteness effects associated with encounters
between particles. Such effects lead to the formation of an isothermal ``core''
of size $\sim r_s$ on the evaporation timescale at the scale radius, $t_{\rm
evap(r_s)}$. At later times, collisional effects lead to the progressive
``collapse'' of this core, increasing the mean density within $1\, r_s$ by
$50\%$ after $t_{\rm evap} (r_s)$. The survival of small-$N$ halos may be
affected by this process in cosmological simulations; in particular, the
evolution of systems with $N \lsim 200$ should be carefully monitored in order
to ensure that discreteness effects do not unduly bias the results.

\item Satellite halos orbiting in the tidal field of a massive host lose mass
continuously as a result of tidal stripping, although the mass loss rate slows
down significantly as the satellite becomes confined within its tidal
radius. When the tidal confinement approaches the minimum binding radius,
$r_{\rm bind}$, NFW models are fully disrupted. Quantitatively, we find that
NFW halos on orbits where $r_t < 2\, r_s \sim 2.6\, r_{\rm bind}$ at pericenter
are fully disrupted.

\item Although tides preferentially strip mass from the outer regions, tides
also cause the halo to expand and the central density to decrease after each
pericentric passage. As a result, simple models based on the tidal-limit
approximation underestimate significantly the total tidal mass loss over several
orbits.

\item The equilibrium structure of stripped NFW halos depends mainly on the
fraction of mass lost, and can be expressed in terms of a simple correction to
the original NFW profile
(eqs.~\ref{eq:rhomod},~\ref{eq:rtembnd},~\ref{eq:ftmbnd}). Models based on the
impulse or tidal approximation that take into account this evolution in
structure are found to account reasonably well for the mass lost through several
pericentric passages.

\end{enumerate}

Applying these results to the dwarf spheroidal companions of the Milky Way, we
conclude that their surrounding dark matter halos have circular velocity curves
that peak well beyond the luminous radius at values significantly higher than
expected from the stellar line-of-sight velocity dispersion. The modeling also
suggests that the true tidal radius of dSphs may lie well beyond the radial
cutoff observed in surface brightness profiles, suggesting that these are not
really tidal in origin but rather are features in the luminous profile of little
dynamical relevance.  Our results for Draco, in particular, strongly suggest
that the tidal radius should be much larger than the $28.3$ arcmin derived by
Irwin \& Hatzidimitriou (1995), in agreement with the recent Sloan survey data
reported by Odenkirchen et al.~(2001).

Our results imply a close correspondence between the most massive substructure
halos and the known satellites of the Milky Way and suggests that only
substructure halos with peak circular velocities below $35$ km s$^{-1}$ lack
readily detectable luminous counterparts. The scarcity of satellites around the
Milky Way just reflects the small number of substructure halos with circular
velocities exceeding $\sim 35 \kms$. This implies that it may be possible to
explain the abundance of dSphs without invoking a substantial role for the
reionization of the universe in hindering the formation of dwarf systems.

Our conclusions also highlight the fact that the properties of galaxies
inhabiting low-mass cold dark matter halos must vary substantially from halo to
halo to be consistent with observations; vastly different galaxies such as the
SMC and Draco are predicted to live in halos of not very dissimilar circular
velocities.  Explaining the diversity of galaxy properties in low-mass halos
remains a challenge to hierarchical models of galaxy formation that is unlikely
to be solved until the effects of mass, collapse epoch, efficiency of feedback,
and environment are properly understood.

\acknowledgements

Lars Hernquist kindly made available software that was modified to generate our
N-body halo models.  We thank Colin Leavett-Brown for expert assistance with
the IBM/SP3 supercomputer at the University of Victoria. We acknowledge many
useful conversations with Simon White. The Natural Sciences \& Engineering
Research Council of Canada (NSERC) and the Canadian Foundation for Innovation
have supported this research through various grants to JFN.

\bibliographystyle{astron}
\bibliography{subs}

\begin{thebibliography}{}

\bibitem[\protect\astroncite{{Aguilar} and {White}}{1985}]{AGU85}
{Aguilar}, L.~A. and {White}, S.~D.~M.: 1985,
\newblock {\em \apj} {\bf 295}, 374

\bibitem[\protect\astroncite{{Avila-Reese} et~al.}{2001}]{AVILA01}
{Avila-Reese}, V., {Col{\' i}n}, P., {Valenzuela}, O., {D'Onghia}, E., and
  {Firmani}, C.: 2001,
\newblock {\em \apj} {\bf 559}, 516

\bibitem[\protect\astroncite{{Bahcall} et~al.}{1999}]{BAHC99}
{Bahcall}, N.~A., {Ostriker}, J.~P., {Perlmutter}, S., and {Steinhardt}, P.~J.:
  1999,
\newblock {\em Science} {\bf 284}, 1481

\bibitem[\protect\astroncite{Benson et~al.}{2001}]{BEN01}
Benson, A.~J., Frenk, C.~S., Lacey, C.~G., Baugh, C.~M., and Cole, S.: 2001,
\newblock Preprint [astro-ph/0108218]

\bibitem[\protect\astroncite{{Bertschinger}}{1985}]{BERT85}
{Bertschinger}, E.: 1985,
\newblock {\em \apjs} {\bf 58}, 39

\bibitem[\protect\astroncite{{Binney} and {Tremaine}}{1987}]{BT87}
{Binney}, J. and {Tremaine}, S.: 1987,
\newblock {\em {Galactic dynamics}},
\newblock Princeton, NJ, Princeton University Press

\bibitem[\protect\astroncite{{Blumenthal} et~al.}{1984}]{BLUM84}
{Blumenthal}, G.~R., {Faber}, S.~M., {Primack}, J.~R., and {Rees}, M.~J.: 1984,
\newblock {\em \nat} {\bf 311}, 517

\bibitem[\protect\astroncite{{Bode} et~al.}{2001}]{BODE01}
{Bode}, P., {Ostriker}, J.~P., and {Turok}, N.: 2001,
\newblock {\em \apj} {\bf 556}, 93

\bibitem[\protect\astroncite{{Bullock} et~al.}{2000}]{BULL00}
{Bullock}, J.~S., {Kravtsov}, A.~V., and {Weinberg}, D.~H.: 2000,
\newblock {\em \apj} {\bf 539}, 517

\bibitem[\protect\astroncite{{Burkert}}{1997}]{BURK97}
{Burkert}, A.: 1997,
\newblock {\em \apjl} {\bf 474}, L99

\bibitem[\protect\astroncite{{Col{\' i}n} et~al.}{2000}]{COLIN00}
{Col{\' i}n}, P., {Avila-Reese}, V., and {Valenzuela}, O.: 2000,
\newblock {\em \apj} {\bf 542}, 622

\bibitem[\protect\astroncite{{Cole} et~al.}{1994}]{COLE94}
{Cole}, S., {Aragon-Salamanca}, A., {Frenk}, C.~S., {Navarro}, J.~F., and
  {Zepf}, S.~E.: 1994,
\newblock {\em \mnras} {\bf 271}, 781

\bibitem[\protect\astroncite{{Colpi} et~al.}{1999}]{COLPI99}
{Colpi}, M., {Mayer}, L., and {Governato}, F.: 1999,
\newblock {\em \apj} {\bf 525}, 720

\bibitem[\protect\astroncite{{Dalcanton} and {Hogan}}{2001}]{DAL01}
{Dalcanton}, J.~J. and {Hogan}, C.~J.: 2001,
\newblock {\em \apj} {\bf 561}, 35

\bibitem[\protect\astroncite{{Davis} et~al.}{1985}]{DAVIS85}
{Davis}, M., {Efstathiou}, G., {Frenk}, C.~S., and {White}, S.~D.~M.: 1985,
\newblock {\em \apj} {\bf 292}, 371

\bibitem[\protect\astroncite{{Dehnen}}{2001}]{DEHNEN01}
{Dehnen}, W.: 2001,
\newblock {\em \mnras} {\bf 324}, 273

\bibitem[\protect\astroncite{{Dubinski} and {Carlberg}}{1991}]{DUB91}
{Dubinski}, J. and {Carlberg}, R.~G.: 1991,
\newblock {\em \apj} {\bf 378}, 496

\bibitem[\protect\astroncite{{Efstathiou}}{1992}]{EFST92}
{Efstathiou}, G.: 1992,
\newblock {\em \mnras} {\bf 256}, 43P

\bibitem[\protect\astroncite{{Eke} et~al.}{2001}]{ENS01}
{Eke}, V.~R., {Navarro}, J.~F., and {Steinmetz}, M.: 2001,
\newblock {\em \apj} {\bf 554}, 114

\bibitem[\protect\astroncite{{Font} et~al.}{2001}]{FONT01}
{Font}, A.~S., {Navarro}, J.~F., {Stadel}, J., and {Quinn}, T.: 2001,
\newblock {\em \apjl} {\bf 563}, L1

\bibitem[\protect\astroncite{{Frenk} et~al.}{1988}]{FREN88}
{Frenk}, C.~S., {White}, S.~D.~M., {Davis}, M., and {Efstathiou}, G.: 1988,
\newblock {\em \apj} {\bf 327}, 507

\bibitem[\protect\astroncite{{Fukushige} and {Makino}}{1997}]{FUK97}
{Fukushige}, T. and {Makino}, J.: 1997,
\newblock {\em \apjl} {\bf 477}, L9

\bibitem[\protect\astroncite{{Ghigna} et~al.}{1998}]{GHIG98}
{Ghigna}, S., {Moore}, B., {Governato}, F., {Lake}, G., {Quinn}, T., and
  {Stadel}, J.: 1998,
\newblock {\em \mnras} {\bf 300}, 146

\bibitem[\protect\astroncite{{Ghigna} et~al.}{2000}]{GHIG00}
{Ghigna}, S., {Moore}, B., {Governato}, F., {Lake}, G., {Quinn}, T., and
  {Stadel}, J.: 2000,
\newblock {\em \apj} {\bf 544}, 616

\bibitem[\protect\astroncite{{Gnedin} et~al.}{1999}]{GNE99}
{Gnedin}, O.~Y., {Lee}, H.~M., and {Ostriker}, J.~P.: 1999,
\newblock {\em \apj} {\bf 522}, 935

\bibitem[\protect\astroncite{{Hernquist}}{1993}]{HERN93}
{Hernquist}, L.: 1993,
\newblock {\em \apjs} {\bf 86}, 389

\bibitem[\protect\astroncite{{Irwin} and {Hatzidimitriou}}{1995}]{IH95}
{Irwin}, M. and {Hatzidimitriou}, D.: 1995,
\newblock {\em \mnras} {\bf 277}, 1354 (IH95)

\bibitem[\protect\astroncite{{Kauffmann} et~al.}{1993}]{KAUFF93}
{Kauffmann}, G., {White}, S.~D.~M., and {Guiderdoni}, B.: 1993,
\newblock {\em \mnras} {\bf 264}, 201

\bibitem[\protect\astroncite{{King}}{1962}]{KING62}
{King}, I.~R.: 1962,
\newblock {\em \aj} {\bf 67}, 471

\bibitem[\protect\astroncite{{King}}{1966}]{KING66}
{King}, I.~R.: 1966,
\newblock {\em \aj} {\bf 71}, 64

\bibitem[\protect\astroncite{{Kleyna} et~al.}{2001}]{KLEYNA01}
{Kleyna}, J.~T., {Wilkinson}, M.~I., {Evans}, N.~W., and {Gilmore}, G.: 2001,
\newblock {\em \apjl} {\bf 563}, L115

\bibitem[\protect\astroncite{{Klypin} et~al.}{1999a}]{KGKK99}
{Klypin}, A., {Gottl{\" o}ber}, S., {Kravtsov}, A.~V., and {Khokhlov}, A.~M.:
  1999a,
\newblock {\em \apj} {\bf 516}, 530 (K99b)

\bibitem[\protect\astroncite{{Klypin} et~al.}{2001}]{KLY01}
{Klypin}, A., {Kravtsov}, A.~V., {Bullock}, J.~S., and {Primack}, J.~R.: 2001,
\newblock {\em \apj} {\bf 554}, 903

\bibitem[\protect\astroncite{{Klypin} et~al.}{1999b}]{KLY99}
{Klypin}, A., {Kravtsov}, A.~V., {Valenzuela}, O., and {Prada}, F.: 1999b,
\newblock {\em \apj} {\bf 522}, 82 (K99a)

\bibitem[\protect\astroncite{{Knebe} et~al.}{2002}]{KNEBE02}
{Knebe}, A., {Devriendt}, J.~E.~G., {Mahmood}, A., and {Silk}, J.: 2002,
\newblock {\em \mnras} {\bf 329}, 813

\bibitem[\protect\astroncite{{Majewski} et~al.}{2000}]{MAJ00}
{Majewski}, S.~R., {Ostheimer}, J.~C., {Patterson}, R.~J., {Kunkel}, W.~E.,
  {Johnston}, K.~V., and {Geisler}, D.: 2000,
\newblock {\em \aj} {\bf 119}, 760

\bibitem[\protect\astroncite{{Mateo}}{1998}]{MATEO98}
{Mateo}, M.~L.: 1998,
\newblock {\em \araa} {\bf 36}, 435

\bibitem[\protect\astroncite{{Mayer} et~al.}{2001}]{MAYER01}
{Mayer}, L., {Governato}, F., {Colpi}, M., {Moore}, B., {Quinn}, T., {Wadsley},
  J., {Stadel}, J., and {Lake}, G.: 2001,
\newblock {\em \apj} {\bf 559}, 754

\bibitem[\protect\astroncite{{Moore} et~al.}{2001}]{MOORE01}
{Moore}, B., {Calc{\' a}neo-Rold{\' a}n}, C., {Stadel}, J., {Quinn}, T.,
  {Lake}, G., {Ghigna}, S., and {Governato}, F.: 2001,
\newblock {\em \prd} {\bf 64}, 063508

\bibitem[\protect\astroncite{{Moore} et~al.}{1999}]{MOORE99}
{Moore}, B., {Ghigna}, S., {Governato}, F., {Lake}, G., {Quinn}, T., {Stadel},
  J., and {Tozzi}, P.: 1999,
\newblock {\em \apjl} {\bf 524}, L19 (M99)

\bibitem[\protect\astroncite{{Moore} et~al.}{1998}]{MOORE98}
{Moore}, B., {Governato}, F., {Quinn}, T., {Stadel}, J., and {Lake}, G.: 1998,
\newblock {\em \apjl} {\bf 499}, L5

\bibitem[\protect\astroncite{{Moore} et~al.}{1996}]{MKL96}
{Moore}, B., {Katz}, N., and {Lake}, G.: 1996,
\newblock {\em \apj} {\bf 457}, 455

\bibitem[\protect\astroncite{Navarro}{2002}]{JFN01}
Navarro, J.~F.: 2002,
\newblock in J. Makino and P.Hut (eds.), {\em Astrophysical SuperComputing
  using Particles}, IAU Symposium No. 208,
\newblock Preprint [astro-ph/0110680]

\bibitem[\protect\astroncite{{Navarro} et~al.}{1996}]{NFW96}
{Navarro}, J.~F., {Frenk}, C.~S., and {White}, S.~D.~M.: 1996,
\newblock {\em \apj} {\bf 462}, 563 (NFW)

\bibitem[\protect\astroncite{{Navarro} et~al.}{1997}]{NFW97}
{Navarro}, J.~F., {Frenk}, C.~S., and {White}, S.~D.~M.: 1997,
\newblock {\em \apj} {\bf 490}, 493

\bibitem[\protect\astroncite{{Odenkirchen} et~al.}{2001}]{ODEN01}
{Odenkirchen}, M., {Grebel}, E.~K., {Harbeck}, D., {Dehnen}, W., {Rix}, H.,
  {Newberg}, H.~J., {Yanny}, B., {Holtzman}, J., {Brinkmann}, J., {Chen}, B.,
  {Csabai}, I., {Hayes}, J.~J.~E., {Hennessy}, G., {Hindsley}, R.~B., {Ivezi{\'
  c}}, {\v Z}., {Kinney}, E.~K., {Kleinman}, S.~J., {Long}, D., {Lupton},
  R.~H., {Neilsen}, E.~H., {Nitta}, A., {Snedden}, S.~A., and {York}, D.~G.:
  2001,
\newblock {\em \aj} {\bf 122}, 2538

\bibitem[\protect\astroncite{{Peebles}}{1984}]{PEE84}
{Peebles}, P.~J.~E.: 1984,
\newblock {\em \apj} {\bf 277}, 470

\bibitem[\protect\astroncite{Power et~al.}{2002}]{POW02}
Power, C., Navarro, J.~F., Jenkins, A., Frenk, C., White, S. D.~M., Springel,
  V., Stadel, J., and Quinn, T.: 2002,
\newblock Preprint [astro-ph/0201544]

\bibitem[\protect\astroncite{{Quinlan}}{1996}]{QUINLAN96}
{Quinlan}, G.~D.: 1996,
\newblock {\em New Astronomy} {\bf 1}, 255

\bibitem[\protect\astroncite{Somerville}{2001}]{SOM01}
Somerville, R.~S.: 2001,
\newblock Preprint [astro-ph/0107507]

\bibitem[\protect\astroncite{{Spergel} and {Steinhardt}}{2000}]{SS00}
{Spergel}, D.~N. and {Steinhardt}, P.~J.: 2000,
\newblock {\em Physical Review Letters} {\bf 84}, 3760

\bibitem[\protect\astroncite{{Springel} et~al.}{2001}]{SPR01}
{Springel}, V., {White}, S.~D.~M., {Tormen}, G., and {Kauffmann}, G.: 2001,
\newblock {\em \mnras} {\bf 328}, 726

\bibitem[\protect\astroncite{Stadel}{2001}]{STADEL01}
Stadel, J.~G.: 2001,
\newblock {\em Ph.D. thesis}, University of Washington

\bibitem[\protect\astroncite{Stoehr et~al.}{2002}]{STOEHR02}
Stoehr, F., White, S. D.~M., Tormen, G., and Springel, V.: 2002,
\newblock Preprint [astro-ph/0203342]

\bibitem[\protect\astroncite{{Taylor} and {Babul}}{2001}]{TAYBAB}
{Taylor}, J.~E. and {Babul}, A.: 2001,
\newblock {\em \apj} {\bf 559}, 716

\bibitem[\protect\astroncite{{Tormen} et~al.}{1998}]{TORM98}
{Tormen}, G., {Diaferio}, A., and {Syer}, D.: 1998,
\newblock {\em \mnras} {\bf 299}, 728

\bibitem[\protect\astroncite{van Kampen}{2000}]{VANK00}
van Kampen, E.: 2000,
\newblock Preprint [astro-ph/0002027]

\bibitem[\protect\astroncite{{Warren} et~al.}{1992}]{WAR92}
{Warren}, M.~S., {Quinn}, P.~J., {Salmon}, J.~K., and {Zurek}, W.~H.: 1992,
\newblock {\em \apj} {\bf 399}, 405

\bibitem[\protect\astroncite{White}{2000}]{WHITE00}
White, S.~D.~M.: 2000,
\newblock {\em ITP Conference on Galaxy Formation and Evolution},
\newblock http://online.itp.ucsb.edu/online/galaxy/c00/white/

\bibitem[\protect\astroncite{{White} and {Rees}}{1978}]{WHITE78}
{White}, S.~D.~M. and {Rees}, M.~J.: 1978,
\newblock {\em \mnras} {\bf 183}, 341

\end{thebibliography}

%\vfil\eject
% Figures at the end of text
%\clearpage

%%%%%%%%%%%%%%%
% Tables 
%%%%%%%%%%%%%%%

\clearpage

%%% TABLE 1 %%%

\begin{deluxetable}{lrrrr}
\tablecaption{Characteristic timescales for an isolated $N=10^6$ NFW Halo\label{tab:1M}}
\tablehead{
\colhead{r} & \colhead{$N_0(r)$} & \colhead{$t_{0}/t_{\rm cross}$} & 
\colhead{$t_{coll}/t_{\rm cross}$} & \colhead{$t_{evap}/t_{\rm cross}$}  \\ 
}

\startdata

         $0.025$        & $100$        & $120$        & $0.17$        & $23$\\
         $0.05$        & $500$        & $190$        & $0.80$        & $110$\\
         $0.1$        & $2500$        & $450$        & $4.0$\pz        & $550$\\
          $0.2$        & $10000$        & $1000$        & $20.0$\pz        & $2640$\\
\enddata

\tablecomments{Collisional relaxation ($t_{\rm relax}$) and evaporation
($t_{\rm evap}$) timescales in the inner regions of an $N=10^6$ NFW halo model.
The timescale for evolution of the inner mass structure observed in the
simulations, $t_0$, is in reasonable agreement with the evaporation timescale,
$t_{\rm evap} \approx 136~ t_{\rm relax}$.}
\end{deluxetable}

%\clearpage

%%% TABLE 2 %%%

\begin {table}
\tabcaption{Parameters of N-body Simulations\label{tab:orbparams}}
\begin{tabular*}{1.0\textwidth}%
     {@{\extracolsep{\fill}}rrrrrrr}
\hline\hline
\\
\multicolumn{7}{c}{{\sc Eccentric Orbits}} \\ \\
$N$ & $M_{\rm host}/m_{\rm sat}$ & $r_{\rm ap}/R_s$ & $r_{\rm per}/R_s$ & $t_{\rm orb}/t_{\rm cross}$ &
$\epsilon_{\rm circ}$ & $r_{t}/r_s$  \\ \\  
\hline \\

$3000,10^5$ & $300$ & $10.0$ & $1.5$ & $4.8$ & $0.50$ & $3.0$ \\
$3000,10^5$ & $300$ & $10.0$ & $3.0$ & $5.4$ & $0.75$ & $4.9$ \\
$3000,10^5$ & $300$ & $10.0$ & $6.0$ & $6.9$ & $0.95$ & $8.8$ \\
\hline\\
\multicolumn{7}{c}{{\sc Circular Orbits}} \\ \\
$N$ & $M_{\rm host}/m_{\rm sat}$ & \multicolumn{2}{c}{$r_{\rm circ}/R_s$} & $t_{\rm orb}/t_{\rm cross}$ &
& $r_{t}/r_s$  \\ \\  
\hline \\

$10^5$ & $3000$ & \multicolumn{2}{c}{$1.5$} &  $0.19$ & &$0.87$ \\
$10^5$ & $1500$ & \multicolumn{2}{c}{$1.5$} &  $0.65$ & &$1.3$\pz \\
$10^5$ & $1000$ & \multicolumn{2}{c}{$1.5$} &  $0.79$ & &$1.6$\pz \\
$10^5$ & $600$  & \multicolumn{2}{c}{$1.5$} &  $1.0$\pz & &$2.1$\pz \\

$3000,10^5$ & $300$ & \multicolumn{2}{c}{$1.5$} &  $1.4$\pz & &$3.0$\pz \\
$3000$ & $300$ & \multicolumn{2}{c}{$3.0$} &  $2.9$\pz & &$4.9$\pz \\
$3000$ & $300$ & \multicolumn{2}{c}{$6.0$} &  $6.2$\pz & &$8.8$\pz \\
$3000$ & $300$ & \multicolumn{2}{c}{$7.5$} &  $8.1$\pz & &$10.6$\pz \\
$3000$ & $300$ & \multicolumn{2}{c}{$10.0$\pz}  & $11.5$\pz & &$13.2$\pz \\
$3000$ & $300$ & \multicolumn{2}{c}{$30.0$\pz}  & $46.3$\pz & &$32.2$\pz \\
\hline \\ \\

\end{tabular*}

\hskip1em{\sc Note. ---} $\epsilon_{\rm circ}= J/J_{\rm
circ}(E)$ is the orbital circularity, defined as the ratio between the
orbital angular momentum and the angular momentum of a circular orbit
with the same energy; $t_{\rm orb}/t_{\rm cross}$ is the orbital
period in units of the crossing time of the satellite halo particles;
and $r_t$ is the estimate of the tidal radius given by
eq.~\ref{eq:rtkly} computed at pericenter.
\end{table}

%%% TABLE 3 %%%

\begin{deluxetable}{lcrcrcrc}
\tablecaption{Structural and kinematic properties of Satellites
of the Milky Way \label{table:dsph}}

\tablehead{
\colhead{Galaxy} & \colhead{${\rm L_{tot}}$ } & \colhead{$r_{\rm core}$ } &
\colhead{$r_{tl}$} & \colhead{$\sigma_*$ } & \colhead{$D$} & \colhead{$z_{acc}$} &
\colhead {$V_{\rm max} (z_{acc})$} \\
       &\colhead{ $(10^6 \,{\rm L_\odot})$} & \colhead{ ${\rm (pc)}$} & \colhead{${\rm
(pc)}$  } & \colhead{$\kms$  } & \colhead{(kpc)} & &\colhead{$\kms$}
}

\startdata
Carina & $0.43$ & $177 \pm 28$ & $581 \pm 86$ & $6.8 \pm
1.6$ & $103 \pm 5$\pz & $2.3$ & $53.1$ \\ 
Draco & $0.26$ & $158 \pm 14$ & $498 \pm 47$ & $9.5 \pm
1.6$ & $82 \pm 5$ & $2.8$ & $73.1$ \\
Fornax & $15.5$ & $400 \pm 43$ & $2078 \pm 177$ & $10.5 \pm
1.5$ & $140 \pm 8$\pz & $1.6$ & $33.0$ \\
Leo I & $4.79$ & $169 \pm 19$ & $645 \pm 87$ & $8.8 \pm
0.9$ & $254 \pm 30$ & $0.6$ & $59.9$ \\ 
Leo II &$0.58$ & $162 \pm 35$ & $487 \pm 60$ & $6.7 \pm
1.1$ & $208 \pm 12$ & $0.9$ & $65.3$ \\
Sculptor & $2.15$ & $101 \pm 28$ & $1329 \pm 107$ & $6.6
\pm 0.7$ & $79 \pm 4$ & $2.9$ & $25.4$ \\
Sextans & $0.50$ & $322 \pm 42$ & \pz$3102 \pm 1028$ & $6.6
\pm 0.7$ & $89 \pm 4$ & $2.6$ & $17.0$ \\
Ursa Minor & $0.29$ & $196 \pm 24$ & $628 \pm 74$ & $9.3
\pm 1.8$ & $68 \pm 3$ & $3.3$ & $93.2$ \\
\enddata

\tablecomments{Column (1) lists the name of the satellite galaxy. Columns
(2), (5) and (6) list the $V$-band luminosity, stellar central velocity
dispersion and galactocentric distance taken from data presented by Mateo
(1998).  Columns (3) and (4) list the King model core radius and tidal radius
derived by Irwin \& Hatzidimitriou (1995).  Columns (7) and (8) list the
estimated accretion redshift (based on the galactocentric distance) and peak
circular velocity of the dark halo surrounding each dwarf spheroidal at that
redshift.  Note that Odenkirchen et al.~(2001) derive much larger values of the
King model core and tidal radii for Draco, $r_{\rm core}=179$ pc, $r_{tl}=1020$
pc.}
\end{deluxetable}

%%%%%%%%%%%%%%%
% Figures Captions
%%%%%%%%%%%%%%%

\clearpage

%%%
\clearpage

\clearpage

\clearpage

\clearpage

\clearpage

\clearpage

\clearpage

\clearpage

\clearpage

\clearpage

\clearpage

\clearpage

\clearpage

\end{document}